\documentclass[10pt,showpacs,floatfix,letterpaper,amsmath,amsfonts,amssymb,aps,pra,reprint,superscriptaddress]{revtex4-1}
\usepackage[latin1]{inputenc}
\usepackage{bm}
\usepackage{graphicx}
\usepackage{tensor}
\usepackage{epstopdf}
\usepackage{tikz}
\usepackage[caption=false]{subfig}

\usetikzlibrary{shapes,positioning}
\tikzset{unit/.style={draw,shape=circle,fill=black,scale=0.2},costate/.style={draw,shape=trapezium},state/.style={draw,shape border rotate=180,shape=trapezium},qo/.style={draw,shape=rectangle},node distance=1mm,scalar/.style={draw,shape=regular polygon,regular polygon sides=6}}

\newcommand{\op}[1]{\hat{#1}}                                 
\newcommand{\rop}[1]{\check{#1}}                              
\newcommand{\qo}[1]{\mathcal{#1}}                             
\newcommand{\ket}[1]{\lvert #1\rangle}                        
\newcommand{\bra}[1]{\langle #1 \rvert}                       
\newcommand{\pr}[1]{\ket{#1}\bra{#1}}                         
\newcommand{\ipr}[2]{\langle #1 ,\, #2 \rangle}               
\newcommand{\mean}[1]{\left\langle #1 \right\rangle}          
\newcommand{\ccmean}[3]{\tensor[_{#1}]{\mean{#2}}{_{#3}}}     
\newcommand{\Tr}[1]{\text{Tr}(#1)}                            

\begin{document}
\title{Quantum instruments as a foundation for both states and observables}
\author{Justin Dressel}
\affiliation{Department of Physics and Astronomy and Rochester Theory Center, University of Rochester, Rochester, New York 14627, USA}
\author{Andrew N. Jordan}
\affiliation{Department of Physics and Astronomy and Rochester Theory Center, University of Rochester, Rochester, New York 14627, USA}
\affiliation{Institute of Quantum Studies, Chapman University, 1 University Drive, Orange, CA 92866, USA}

\date{\today}

\begin{abstract}
  We demonstrate that quantum instruments can provide a unified operational foundation for quantum theory.  Since these instruments directly correspond to laboratory devices, this foundation provides an alternate, more experimentally grounded, perspective from which to understand the elements of the traditional approach.  We first show that in principle all measurable probabilities and correlations can be expressed entirely in terms of quantum instruments without the need for conventional quantum states or observables.  We then show how these states and observables reappear as derived quantities by conditioning joint detection probabilities on the first or last measurement in a sequence as a preparation or a post-selection.  Both predictive and retrodictive versions of states and observables appear in this manner, as well as more exotic bidirectional and interdictive states and observables that cannot be easily expressed using the traditional approach.  We also revisit the conceptual meaning of the Heisenberg and Schr\"{o}dinger pictures of time evolution as applied to the various derived quantities, illustrate how detector loss can be included naturally, and discuss how the instrumental approach fully generalizes the time-symmetric two-vector approach of Aharonov \emph{et al.} to any realistic laboratory situation.
\end{abstract}

\maketitle


\section{Introduction}
Hiding beneath the conceptual trappings of modern quantum mechanics \cite{Dirac1930,VonNeumann1932,Alicki2001,Wiseman2009} lies an inference formalism.  Like probability theory \cite{Jaynes2003}, this formalism provides a self-consistent logic for manipulating uncertainty about Boolean (true/false) propositions.  Unlike standard probability theory, this formalism describes collections of propositions that may not be mutually exclusive \cite{Jauch1968,Dressel2012b,Dressel2013}, so cannot be simultaneously tested or combined with the logical operations of \textsc{and}/\textsc{or}.  An experimenter typically uses this inference formalism to \emph{predict} the likelihoods that future measurement events will occur on macroscopic laboratory instruments---such as avalanche photo-diodes, spectrometers, or scintillators---given that some repeatable preparation event has occurred.  The information about the preparation procedure that is needed to make predictions of this sort is encoded into a mathematical object known as the \emph{predictive quantum state}.

An inference formalism need not be used only to make predictions, however.  Indeed, as early as 1955 Watanabe observed that it was equally possible to retroactively infer---or \emph{retrodict}---the likelihoods that specific preparation events had occurred if one knew which posterior events were later measured \cite{Watanabe1955}.  This observation led to the definition of the \emph{retrodictive quantum state}, which, analogously to the predictive quantum state, is a mathematical object that encodes the information about a posterior measurement event needed to make retrodictions of this sort.  This alternate approach to the quantum formalism was rediscovered by Aharonov \emph{et al.} in 1964 \cite{Aharonov1964}, and again in 1999 by Pegg and Barnett \cite{Pegg1999}, which has since prompted theoretical development by many others \cite{Barnett2000a,Pegg2002a,Pegg2002b,Chefles2003,Hofmann2003,Pregnell2004,Vaidman2007,Pegg2008,Scroggie2008,Amri2011}.  In the experimental realm the retrodictive state has been successfully used to describe atom-photon emission \cite{Barnett2000b,Barnett2000c,Barnett2001,Jeffers2002a}, phase measurements \cite{Pregnell2001,Pegg2005,Resch2007}, field measurements \cite{Jeffers2002b,Tan2004,Ban2007}, and optical state engineering \cite{Jedrkiewicz2004}.

During this most recent development period for the retrodictive quantum state, there have also been parallel efforts to more explicitly recast quantum mechanics as an inference formalism that generalizes Bayesian probability theory.  Korotkov and ANJ have shown that Bayesian inference can correctly predict the outcomes of continuous quantum measurements \cite{Korotkov1999,Korotkov2006,Jordan2006,Williams2008}.  Caves, Fuchs, Spekkens, Harrigan, and Bartlett have all proposed a Bayesian interpretation for the predictive quantum state \cite{Caves2002,Spekkens2007,Harrigan2010,Bartlett2012}.  The present authors developed an algebraic approach to Bayesian probability theory that can express both classical and quantum measurements using the same language \cite{Dressel2012b,Dressel2013}.  Leifer and Spekkens have explicitly constructed a causally neutral theory of noncommutative Bayesian inference using conditional quantum states that serve as generalizations of conditional probabilities \cite{Leifer2006,Leifer2007,Leifer2008,Leifer2011}.  Abramsky \cite{Abramsky2004}, Selinger \cite{Selinger2004,Selinger2007}, and Coecke \cite{Coecke2012} have even used abstract category theory to identify the common structural foundations of quantum mechanics and Bayesian probability theory, which has enabled rigorous graphical proofs of quantum information theorems. 

In this work, we supplement these efforts by pursuing a related line of thought inspired by an observation recently made by Rau \cite{Rau2011}: quantum theory is the unique extension to probability theory that can emulate all evolution by sequences of measurements.  In light of this observation and the fact that only measurement events possess any defensibly real status in the laboratory, we argue that we should consider expressing quantum theory entirely in terms of these measurement events.  Surprisingly, we show that such a conceptual reformulation is possible---predictive states, retrodictive states, and even quantum observables can be treated as \emph{derived} quantities from a single mathematical entity that directly corresponds to a laboratory detector: the quantum instrument \cite{Davies1970,Ozawa1984}.  The resulting formulation is effectively \emph{stateless}, which adds a new spin to the continued controversy regarding the significance of the quantum state.

We emphasize that in addition to reproducing existing results in the literature, this stateless reformulation organically extends to underexplored territory.  For example, interdictive states and bidirectional states can be derived in addition to the standard predictive and retrodictive states---these more exotic types of state cannot be easily expressed as single density operators.  Similarly, their associated interdictive and bidirectional observables cannot be easily expressed as single Hermitian operators.  The interdictive state is a qualitatively new object to our knowledge, but may have natural applications for eavesdropping scenarios in quantum information protocols.  The bidirectional state fully generalizes the two-vector formalism of Aharonov \emph{et al.} \cite{Aharonov1988,Duck1989,Aharonov2008,Aharonov2009,Aharonov2010} and closely connects to work by Crutchfield \emph{et al.} in characterizing classical stochastic processes \cite{Crutchfield2009,Ellison2009,Ellison2011}.  The impact of detector loss can be automatically included in all these scenarios.  

This paper is organized as follows.  In Section~\ref{sec:instruments} we review the definition of a quantum instrument and give an illustrative example.  In Section~\ref{sec:observables} we show how instruments subsume and generalize the concept of measurable observables.  In Section~\ref{sec:nostate} we introduce our stateless reformulation.  In Section~\ref{sec:retroobs} we show how instruments also generate retrodictive observables.  In Section~\ref{sec:states} we show how predictive states, retrodictive states, interdictive states, and bidirectional states will naturally (re)appear from the stateless formulation with different choices of conditioning.  In Section~\ref{sec:restate} we reintroduce states into the stateless formulation for pragmatic completeness.  We conclude in Section~\ref{sec:conclusion}.

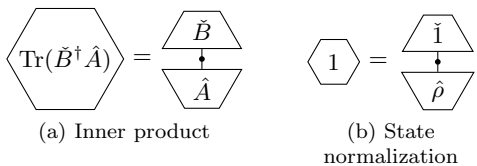
\begin{figure}
  \begin{center}
  \subfloat[Inner product]{
    \begin{tikzpicture}
      \node[scalar,inner sep=-4] (IP) {$\Tr{\rop{B}^\dagger \op{A}}$};
      \node[unit] (t) [node distance=1cm,right=of IP] {};
      \node [right=0mm of IP] {=};
      \node[costate] (FS) [above=of t] {$\rop{B}$} edge [-] (t);
      \node[state] (IS) [below=of t] {$\op{A}$} edge [-] (t);
    \end{tikzpicture}
    }
    \qquad
  \subfloat[State normalization]{
    \begin{tikzpicture}
      \node[scalar] (IP) {1};
      \node[unit] (t) [right=1cm of IP] {};
      \node[right=0mm of IP] {=};
      \node[costate] (F) [above=of t] {$\rop{1}$} edge [-] (t);
      \node[state] (I) [below=of t] {$\op{\rho}$} edge [-] (t);
    \end{tikzpicture}
  }
\end{center}
\caption{(a) Graphical depiction of the Hilbert-Schmidt inner product $\ipr{\rop{B}}{\op{A}}$.  The scalar value of $\Tr{\rop{B}^\dagger \op{A}}$ (hexagon) is conceptually separated into complementary halves $\rop{B}$ (trapezoid) and $\op{A}$ (inverse trapezoid). The choice of hat on the operators indicates this distinction that is induced by the inner product.  (b) The inner product being used to show the normalization of a quantum state $\op{\rho}$.}
  \label{fig:ipr}
\end{figure}
  
\section{Quantum instruments}\label{sec:instruments}

The idea of a quantum instrument (QI) was introduced to the quantum information community in 1970 by Davies and Lewis \cite{Davies1970} and was later refined by Ozawa in 1984 \cite{Ozawa1984}.  Physically, it constitutes the most complete description of the operation of a laboratory detector that possesses a set of distinguishable outcomes.  Each distinguishable combination of outcomes for the detector corresponds to a particular transformation that specifies how observing those outcomes will affect future observations made by other detectors.  Moreover, each outcome can be freely labeled by an experimenter to extract specific averaged information from the measurement in post-processing.

Mathematically, we define a QI initially in terms of predictive quantum states, keeping in mind that we will revisit the role of these states later.  Recall that a quantum state $\rho$ is generally defined as a positive probability functional over a noncommutative enveloping algebra for a continuous group of symmetries \cite{Alicki2001,Dressel2013}.  For simplicity, we assume that this algebra can be represented as an operator algebra over an auxiliary Hilbert space, as is standard practice.  

For later convenience, we express the action of a state functional $\rho$ on an arbitrary operator $\rop{O}$ as an inner product $\rho(\rop{O}) = \ipr{\rop{O}}{\op{\rho}}$ with a positive trace-density operator $\op{\rho}$ in the operator algebra itself; the inner product we use is the Hilbert-Schmidt inner product
\begin{align}\label{eq:ipr}
  \ipr{\rop{B}}{\op{A}} = \Tr{\rop{B}^\dagger \op{A}}.
\end{align}
States are normalized by their action on the unit operator $\rho(\rop{1}) = \ipr{\rop{1}}{\op{\rho}} = 1$.  Note that we denote Hilbert space operators (other than $\op{\rho}$ and $\rop{1}$) in upper case Roman font. We may also notate either hats or inverted hats on the operators to indicate conceptual differences stemming from the directionality of the complex inner product of Eq.~\eqref{eq:ipr}.  The significance of these purely notational distinctions will become clear as our discussion develops.  We illustrate the role of the inner product in Figure~\ref{fig:ipr} and throughout this work, where we have deliberately chosen a graphical form similar to the related category theory work \cite{Coecke2012}.  

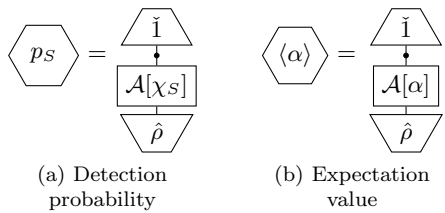
\begin{figure}
  \subfloat[Detection probability]{
    \begin{tikzpicture}[scale=.4]
      \node[scalar] (IP) {$p_S$};
      \node[right=0mm of IP] {=};
      \node[unit] (t) [right=1cm of IP] {};
      \node[qo] (A) [below=of t] {$\qo{A}[\chi_S]$} edge [-] (t);
      \node[costate] (F) [above=of t] {$\rop{1}$} edge [-] (t);
      \node[state] (I) [below=of A] {$\op{\rho}$} edge [-] (A);
    \end{tikzpicture}
    \label{fig:deta}
  }
  \qquad
  \subfloat[Expectation value]{
    \begin{tikzpicture}[scale=.4]
      \node[scalar,inner sep=1] (IP) {$\mean{\alpha}$};
      \node[right=0mm of IP] {=};
      \node[unit] (t) [right=1cm of IP] {};
      \node[qo] (A) [below=of t] {$\qo{A}[\alpha]$} edge [-] (t);
      \node[costate] (F) [above=of t] {$\rop{1}$} edge [-] (t);
      \node[state] (I) [below=of A] {$\op{\rho}$} edge [-] (A);
    \end{tikzpicture}
    \label{fig:detb}
  }
  \caption{(a) The quantum instrument $\qo{A}$ representing a detector is a transformation-valued functional that produces quantum operations $\qo{A}[\chi_S]$ corresponding to each set of detector outcomes $S$, as in Eq.~\eqref{eq:qoS}.  Here $\chi_S$ is a suitable indicator function for the set $S$.  Computing the modified norm using the inner product yields the probability $p_S$ for detecting the outcomes $S$.  (b) More generally, by assigning an appropriate set of values $\alpha$ to each detector outcome, the same quantum instrument can be used to compute any expectation value $\mean{\alpha}$ that can be measured by the detector.}
  \label{fig:statenorm}
\end{figure}
  
A QI that represents a detector is a collection of \emph{transformations}, known as \emph{quantum operations} (QO).  We will use calligraphic font to distinguish these transformations from operators.  Formally, the QI is a QO-valued \emph{measure} $\textrm{d}\qo{A}$ over the set of outcomes $X$ of the detector.  Each QO assigned by the measure is a completely positive map (or super-operator) \cite{Alicki2001,Nielsen2000,Wiseman2009}.  The action of such a map on an arbitrary operator $\op{O}$ can be written in an operator-sum form
\begin{align}\label{eq:qo}
  \textrm{d}\qo{A}(x)\op{O} = \textrm{d}x\, \int_Y \! M_{x,y} \op{O} M_{x,y}^\dagger \, \textrm{d}y 
\end{align}
in terms of ``sandwich'' products with a (generally non-unique) collection of \emph{Kraus} \cite{Kraus1971} or \emph{measurement operators} $M_{x,y}$, for which we omit hats.  Most laboratory detector outcomes can be expressed with either a single measurement operator or a discrete sum, but we keep the notation general here for emphasis.

When such a QO is applied to a state $\op{\rho}$, it will transform it to a new state $\op{\rho}'_x$ scaled by a \emph{probability measure} $\textrm{d}p(x)$ that indicates the likelihood of detecting the outcomes $x\in X$: 
\begin{align}\label{eq:qoupdate}
  \textrm{d}\qo{A}(x)\op{\rho} = \textrm{d}p(x)\, \op{\rho}'_x.
\end{align}
This probability measure can be extracted by computing the modified norm using the inner product, $\ipr{\rop{1}}{\textrm{d}\qo{A}(x)\op{\rho}} = \textrm{d}p(x)$.  It follows that the detector QOs must satisfy a normalization condition $\int_X \ipr{\rop{1}}{\textrm{d}\qo{A}(x)\op{\rho}} \leq 1$ in addition being completely positive; when the detector has no loss, the equality holds and we call the detector \emph{complete}.

An experimenter can freely assign numerical labels to each outcome of the detector, which will specify a function $\alpha(x)$ to be integrated with the QI measure.  Hence, one can also understand a detector QI as a transformation-valued \emph{functional}
\begin{align}\label{eq:qi}
  \qo{A}[\alpha] &= \int_X \alpha(x)\, \textrm{d}\qo{A}(x).
\end{align}
Most laboratory detectors have a finite number of discrete outcomes, so these integration measures will typically reduce to a finite sum.

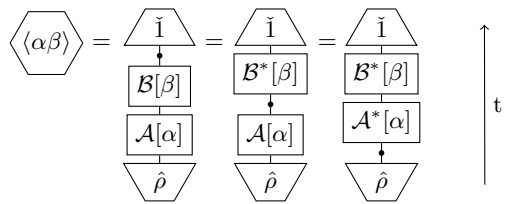
\begin{figure}
    \begin{tikzpicture}[scale=.4]
      \node[scalar,inner sep=-1] (IP) {$\mean{\alpha\beta}$};
      \node[right=0mm of IP] {=};
      \node[costate] (F) [right=8mm of IP.30] {$\rop{1}$};
      \node[unit] (t) [below=of F] {} edge [-] (F);
      \node[qo] (B) [below=of t] {$\qo{B}[\beta]$} edge [-] (t);
      \node[qo] (A) [below=of B] {$\qo{A}[\alpha]$} edge [-] (B);
      \node[state] (I) [below=of A] {$\op{\rho}$} edge [-] (A);
      \node[right=1.5cm of IP] {=};
      \node[costate] (F2) [right=8mm of F] {$\rop{1}$};
      \node[qo] (B2) [below=of F2] {$\qo{B}^*[\beta]$} edge [-] (F2);
      \node[unit] (t2) [below=of B2] {} edge [-] (B2);
      \node[qo] (A2) [below=of t2] {$\qo{A}[\alpha]$} edge [-] (t2);
      \node[state] (I2) [below=of A2] {$\op{\rho}$} edge [-] (A2);
      \node[right=3cm of IP] {=};
      \node[costate] (F3) [right=8mm of F2] {$\rop{1}$};
      \node[qo] (B3) [below=of F3] {$\qo{B}^*[\beta]$} edge [-] (F3);
      \node[qo] (A3) [below=of B3] {$\qo{A}^*[\alpha]$} edge [-] (B3);
      \node[unit] (t3) [below=of A3] {} edge [-] (A3);
      \node[state] (I3) [below=of t3] {$\op{\rho}$} edge [-] (t3);
      \node (t0) [right=1cm of I3.330] { };
      \node (tf) [right=1cm of F3.30] { };
      \draw[->] (t0) to node [right] {t} (tf);
    \end{tikzpicture}
    \caption{The measurable correlation $\mean{\alpha \beta}$ as in Eq.~\eqref{eq:twoqi} between the outcomes of two detectors $\qo{A}$ and $\qo{B}$ arranged in a sequence.  We show this correlation represented three different ways as in Eq.~\eqref{eq:sequence} using the adjoint instruments $\qo{A}^*$ and $\qo{B}^*$. Each diagram corresponds to a different choice of temporal reference point---indicated by the black dot---that conceptually separates future detections from past detections in accordance with the timeline on the right.}
    \label{fig:sequence}
\end{figure}
  
If the labels $\alpha(x)$ are chosen to be indicator functions $\chi_S(x)$ with a value of $1$ for any $x$ in some subset $S\subset X$ of detector outcomes and $0$ otherwise, then the detector QI outputs the appropriate QO that course-grains those detector outcomes
\begin{align}\label{eq:qoS}
  \qo{A}_S = \qo{A}[\chi_S] = \int_X \chi_S(x)\, \textrm{d}\qo{A}(x) = \int_S \textrm{d}\qo{A}.
\end{align}
Computing the modified norm yields the probability $\ipr{\rop{1}}{\qo{A}_S\op{\rho}} = p_S$ for detecting the outcomes in the set $S$, as shown in Figure~\ref{fig:deta}. The maximally course-grained QO associated with a QI is its \emph{non-selective} measurement $\qo{A} = \qo{A}_X = \qo{A}[1] = \int_X \textrm{d}\qo{A}$ that does not discriminate between any of the outcomes of the detector.

The utility of a QI is not restricted to producing QO, however.  It can also produce new types of (non-positive) transformations that depend on the choice of labeling function $\alpha(x)$.  Applying such a transformation to a state produces a weighted sum of modified states, $\qo{A}[\alpha]\op{\rho} = \int \alpha(x)\, \textrm{d}p(x)\, \op{\rho}'_x$, which has a modified norm equal to an average $\ipr{\rop{1}}{\qo{A}[\alpha]\op{\rho}} = \int \alpha(x)\, \textrm{d}p(x) = \mean{\alpha}$ of the labeling function $\alpha$, as indicated in Figure~\ref{fig:detb}.  This average $\mean{\alpha}$ is precisely the statistical mean of the values $\alpha(x)$ that would be reported by an experimenter after recording a large number of measurements by the detector prepared with the state $\op{\rho}$.  

Most importantly, the QIs for a sequence of detectors, such as $\qo{A}$ and $\qo{B}$, may be \emph{composed} to compute the experimentally accessible averages of their joint outcomes: 
\begin{align}\label{eq:twoqi}
  \mean{\alpha\beta} &= \ipr{\rop{1}}{\qo{B}[\beta]\qo{A}[\alpha]\op{\rho}} \\
  &= \int_{X_1}\alpha(x_1)\,\ipr{\rop{1}}{\qo{B}[\beta]\op{\rho}'_{x_1}}\, \mathrm{d}p_1(x_1) \nonumber \\
  &= \int_{X_2}\int_{X_1}\alpha(x_1)\,\beta(x_2)\,\mathrm{d}p_2(x_2|x_1)\,\mathrm{d}p_1(x_1),  \nonumber
\end{align}
as shown in Figure~\ref{fig:sequence}.  Generally, the joint probability measure for the succession of measurements $\mathrm{d}p_2(x_2|x_1)\mathrm{d}p_1(x_1)$ will be correlated.  The transformative nature of each QI is essential for correctly computing these measurable correlations.

For each QI we also define an \emph{adjoint} QI using the inner product of Eq.~\eqref{eq:ipr}, which will be useful in the discussion to follow.  For example, we can rewrite the joint correlation in Eq.~\eqref{eq:twoqi} in several ways 
\begin{align}\label{eq:sequence}
  \ipr{\rop{1}}{\qo{B}[\beta]\qo{A}[\alpha]\op{\rho}} &= \ipr{\qo{B}^*[\beta]\rop{1}}{\qo{A}[\alpha]\op{\rho}} \\
  &= \ipr{\qo{A}^*[\alpha]\qo{B}^*[\beta]\rop{1}}{\op{\rho}} \nonumber
\end{align}
in terms of the adjoint QIs $\qo{A}^*[\alpha]$ and $\qo{B}^*[\beta]$ composed of adjoint measures of the form
\begin{align}
  \textrm{d}\qo{A}^*(x)\rop{O} = \textrm{d}x\, \int_Y M^\dagger_{x,y} \rop{O} M_{x,y} \, \textrm{d}y,
\end{align}
shown acting on an arbitrary operator $\rop{O}$.  These measures differ from the form in Eq.~\eqref{eq:qo} only by the inverted order of the sandwich product, which follows from the cyclic property of the trace in the inner product of Eq.~\eqref{eq:ipr}.  The different ways of writing the correlations using the adjoint QIs correspond to different choices of the conceptual split between a future and a past within the bracketed time interval for the sequence of detections, as indicated in Figure~\ref{fig:sequence}.

\subsection{Example: time evolution}
A simple and trivial illustration of a QI is a unitary time evolution channel $\qo{U}_t$, known as a \emph{propagator}, that indicates a transformation occurring between detections.  In the laboratory one can envision such a channel as a connecting element---such as free space, or an optical fiber---that performs no filtering measurement by itself but does influence the likelihoods of subsequent detections. 

The QI for such a channel has a single possible outcome, so is also a single QO
\begin{align}\label{eq:schrod}
  \qo{U}_t \op{\rho} = U_t \op{\rho} U_t^\dagger
\end{align}
that produces an updated state with probability $1$.  One could assign a label to this single outcome, but we omit it here.  This QO is composed of a single sandwich product with a unitary Kraus operator $U_t$ and corresponds to the Schr\"{o}dinger picture of time evolution.  Its adjoint corresponds to the Heisenberg picture of time evolution
\begin{align}\label{eq:heisen}
  \qo{U}^*_t \rop{O} = U_t^\dagger \rop{O} U_t,
\end{align}
where $\rop{O}$ is an arbitrary operator. The Kraus operator in both cases is the exponentiation $U_t = \exp(t H/i\hbar)$ of a generating Hamiltonian operator $H$ over a time interval $t$.  This Hamiltonian characterizes the symmetry constraints of the propagation.

Since time evolution is an already familiar and well-studied special case of a QI, we will omit it as implicit in the discussion to follow in order to focus on clarifying other aspects of QIs.  In practice, unitary channels will appear between most detecting elements due to evolution inside the connecting regions.  We can thus imagine detector QIs to contain implicit compositions of QIs and unitary connections.  Other types of unitary evolution that are not parametrized by time---such as the effect of a half-wave plate on an optical beam---will also correspond to similar trivial QIs that serve as connecting elements.

\subsection{Example: photodetector}
As a simple but nontrivial illustration of a QI, let us consider an ideal number-resolving photodetector that can identify the total number (including zero) of detected photons of a particular frequency $\omega$ up to a maximum collected number $N$, after which the detector saturates.  Such a detector will have $N+1$ distinguishable outcomes corresponding to the different absorption numbers.  The outcomes from $n=0$ to $n=N-1$ will indicate a definite collection of a particular number of photons, so will have measurement operators of the form $M_n = \ket{0}\bra{n}$. That is, $n$ photons will be absorbed to leave zero remaining detectable photons.  The final outcome for $n=N$ will register for $N$ or greater numbers of collected photons, so it will involve a sum of an infinite number of similar measurement operators.  

Following Eqs. \eqref{eq:qo} and \eqref{eq:qi}, the QI for this detector, which we denote as $\qo{P}$ throughout the paper, can be written as
\begin{align}\label{eq:photo}
  \qo{P}[\alpha]\op{\rho} &= \sum_{n=0}^{N-1} \alpha_n \ket{0}\bra{n}\op{\rho}\ket{n}\bra{0} + \alpha_N \sum_{k=N}^\infty \ket{0}\bra{k}\op{\rho}\ket{k}\bra{0}, \nonumber \\
  &= \pr{0} \left[ \sum_{n=0}^{N-1} \alpha_n p_n + \alpha_N \sum_{k=N}^\infty p_k \right],
\end{align}
where each $\alpha_n$ is a detector label for the outcome $n$, and where we have noted that $p_n = \bra{n}\op{\rho}\ket{n}$ is the probability for detecting $n$ photons given the preparation $\op{\rho}$.  As expected, the updated preparation for any outcome is $\pr{0}$ since all available photons will be collected, leaving the vacuum behind.

Choosing different labels $\alpha_n$ allows the QI in Eq.~\eqref{eq:photo} to compute any quantity that is measurable with this detector.  For example, choosing a single $\alpha_k = 1$ with the rest $0$ will compute the probability $p_k$ for detecting $k$ photons.  Alternatively, choosing $\alpha_n = E_n = n\hbar\omega$ will compute the average resolvable photon energy biased by the saturation of the detector.  These computations will all be encoded into the modified norm $\ipr{\rop{1}}{\qo{P}[\alpha]\op{\rho}}$, which can also be written as $\ipr{\qo{P}^*[\alpha]\rop{1}}{\op{\rho}}$ in terms of the adjoint QI of the detector
\begin{align}\label{eq:photoad}
  \qo{P}^*[\alpha]\rop{O} &= \sum_{n=0}^{N-1} \alpha_n \ket{n}\bra{0}\rop{O}\ket{0}\bra{n} + \alpha_N \sum_{k=N}^\infty \ket{k}\bra{0}\rop{O}\ket{0}\bra{k} \nonumber \\
  &= \bra{0}\rop{O}\ket{0} \left[ \sum_{n=0}^{N-1} \alpha_n \pr{n} + \alpha_N \sum_{k=N}^\infty \pr{k} \right].
\end{align}
This adjoint has a rather different form from Eq.~\eqref{eq:photo}, and is shown here acting on an arbitrary operator $\rop{O}$.

The detector QIs in Eqs.~\eqref{eq:photo} and \eqref{eq:photoad} can be used to compute conditional quantities as well by renormalizing the total detected probability.  For example, to determine the average collected energy for events that do \emph{not} saturate the detector, we first compute the total probability for those detections, $q = \ipr{\rop{1}}{\qo{P}[\chi_q]\op{\rho}} = \ipr{\qo{P}^*[\chi_q]\rop{1}}{\op{\rho}} = \sum_{n=0}^{N-1} p_n$, with an indicator function $\chi_q$ that is $0$ only for the saturated outcome $n=N$, and $1$ for the rest.  We then renormalize the average energy by excluding the outcome $N$ to find the conditioned energy average
\begin{align}\label{eq:condenergy}
  \mean{E}_{n\neq N} = \frac{\ipr{\rop{1}}{\qo{P}[E \chi_q]\op{\rho}}}{\ipr{\rop{1}}{\qo{P}[\chi_q]\op{\rho}}} = \sum_{n=0}^{N-1} E_n \frac{p_n}{q}.
\end{align}
Each $p_n/q$ is the proper conditional probability for obtaining outcome $n$ when the saturated outcome is discarded.

\section{Predictive observables}\label{sec:observables}
\begin{figure}
  \subfloat[Probability operator]{
  \begin{tikzpicture}[scale=.4]
    \node[unit] (IP) {};
    \node[qo] (A) [above=of IP] {$\qo{A}^*[\chi_S]$} edge [-] (IP);
    \node[costate] (F1) [above=of A] {$\rop{1}$} edge [-] (A);
    \node[right=7mm of IP] {=};
    \node[unit] (t) [right=20mm of IP] {};
    \node[costate] (F) [above= of t] {$\rop{A}[\chi_S]$} edge [-] (t);
  \end{tikzpicture}
  \label{fig:po}
  }
  \qquad
  \subfloat[Observable operator]{
  \begin{tikzpicture}[scale=.4]
    \node[unit] (IP) {};
    \node[qo] (A) [above=of IP] {$\qo{A}^*[\alpha]$} edge [-] (IP);
    \node[costate] (F1) [above=of A] {$\rop{1}$} edge [-] (A);
    \node[right=7mm of IP] {=};
    \node[unit] (t) [right=20mm of IP] {};
    \node[costate] (F) [above= of t] {$\rop{A}[\alpha]$} edge [-] (t);
  \end{tikzpicture}
  \label{fig:predobs}
  }
  \caption{Predictive observable operators.  (a) The adjoint quantum instrument $\qo{A}^*$ for the final measured detector generates predictive probability operators $\rop{A}[\chi_S]$ in a POM from indicator functions $\chi_S$ for sets $S$ of detector outcomes, as in Eq.~\eqref{eq:po}.  (b) Similarly, weighting the adjoint instrument outcomes with real contextual values $\alpha$ produces traditional predictive (Hermitian) observable operators $\rop{A}[\alpha]$ that can be indirectly measured by the final detector, as in Eq.~\eqref{eq:pom}.}
  \label{fig:observables}
\end{figure}
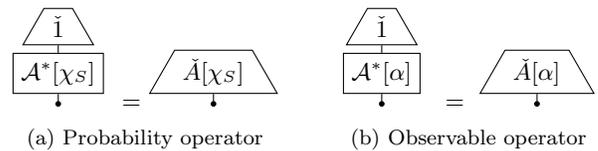

The standard notion of a predictive quantum observable operator can be recovered from a QI provided that we only consider the final detector to be measured.  To see this, we first observe that an adjoint QO applied to the identity produces a positive probability operator (PO)
\begin{align}\label{eq:po}
  \qo{A}^*_S\rop{1} = \rop{A}_S = \int_S\text{d}x\int_Y\text{d}y\, M^\dagger_{x,y}M_{x,y},
\end{align}
as shown in Figure~\ref{fig:po}.  We thus recover the detection probability formula mandated by Gleason's theorem \cite{Gleason1957}
\begin{align}
  p_S = \ipr{\rop{1}}{\qo{A}_S\op{\rho}} = \ipr{\rop{A}_S}{\op{\rho}} = \Tr{\rop{A}_S\op{\rho}}.
\end{align}
Notice that we distinguish PO by inverted hats since they are defined from the action of an adjoint QO.

It follows that a QI for a single measurement will induce a corresponding \emph{probability-operator measure} (POM) \footnote{A POM also has the common name of positive-operator-valued measure (POVM).} according to 
\begin{align}\label{eq:pom}
  \qo{A}^*[\alpha]\rop{1} &= \rop{A}[\alpha] = \int_X \alpha(x)\, \textrm{d}\rop{A}_x.
\end{align}
This POM $\rop{A}[\alpha]$ is an operator-valued \emph{functional} of the labels $\alpha(x)$.  If the QI is complete---so all possible outcomes of the detector are accounted for---then its generated detection probabilities sum to 1 and its associated POM will form a partition of unity $\rop{A}[1] = \rop{1}$.  However, we also allow for the possibility of incomplete QI that can account for inaccessible (loss) outcomes.

The quantity $\rop{A}[\alpha]$ produced from a POM in \eqref{eq:pom} must be a Hermitian operator if the chosen labels $\alpha(x)$ are real, so will be an \emph{observable} in the usual quantum mechanical sense, as shown in Figure~\ref{fig:predobs}.  Furthermore, we see that the statistical average $\mean{\alpha} = \int \alpha(x)\,\textrm{d}p(x) = \Tr{\rop{A}[\alpha]\op{\rho}}$ of the detector labels $\alpha(x)$ produces the standard expectation value of the observable $\rop{A}[\alpha]$.  

Since Eq.~\eqref{eq:pom} need not be the spectral expansion for the observable $\rop{A}[\alpha]$, the chosen detector labels $\alpha(x)$ act as a \emph{generalized spectrum} for the observable.  We dubbed these generalized spectra the \emph{contextual values} for the observable in previous work \cite{Dressel2010,Dressel2012b,Dressel2013}, since the values characterizing the observable depend on the context of the detector being used.  Indeed, a different detecting QI, such as $\qo{B}$, can be used to measure the same observable $\rop{A}[\alpha]$ if appropriately matching values $\beta$ can be found to ensure that $\qo{B}^*[\beta]\rop{1} = \rop{B}[\beta] = \rop{A}[\alpha]$.  In this sense, a Hermitian observable operator represents an \emph{equivalence class} of possible detection strategies for the same average information.

To illustrate how a POM differs from a QI, let us revisit the photodetector example.  Its POM has the form
\begin{align}\label{eq:photopom}
  \rop{P}[\alpha] &= \qo{P}^*[\alpha]\rop{1} = \sum_{n=0}^{N-1} \alpha_n \pr{n} + \alpha_N \sum_{k=N}^\infty \pr{k},
\end{align}
according to Eqs.~\eqref{eq:photoad} and \eqref{eq:po}, and contains only the projections $\pr{n}$ onto specific photon numbers.  Thus, the photodetector POM also happens to be a \emph{projection-valued measure} (PVM).  It partitions unity $\rop{P}[1] = \rop{1}$, which indicates a complete measurement.  Assigning the labels $\alpha_n = E_n = n \hbar\omega$ as before constructs a Hermitian energy observable that conforms to the saturation bias of the detector.  Similarly, assigning a characteristic function that isolates one outcome $k$ will construct a projector $\pr{k}$ as the measured observable; however, just because the measured observable is a projector does \emph{not} imply that the detector prepares a state $\pr{k}$ for subsequent detections. The correct transformation information in the QI of Eq.~\eqref{eq:photo} has been lost by restricting its description to the POM in Eq.~\eqref{eq:photopom}.  

Evidently, this loss of information makes a POM and its generated observables inadequate for computing correlations between sequences of measurements.  It will only provide the same information as a QI when the corresponding detector is the \emph{final} detector to be measured in a sequence of detections.  

To emphasize this point, let us consider a PO for two consecutive measurements, such as those in Eq.~\eqref{eq:twoqi}.  Only the PO for the last measurement $\qo{B}^*_{S_2}\rop{1} = \rop{B}_{S_2}$ will appear in the joint PO
\begin{align}
  \qo{A}^*_{S_1}\qo{B}^*_{S_2}\rop{1} &= \int_{S_1}\textrm{d}x_1\int_Y M^\dagger_{x_1,y}\rop{B}_{S_2}M_{x_1,y},
\end{align}
which cannot be constructed solely from the associated POs $\rop{A}_{S_1}$ and $\rop{B}_{S_2}$ of the two measurements.  Indeed, the full adjoint QI $\qo{A}^*$ for the first measurement is necessary to construct the joint POM for the sequence of measurements.  We must conclude that a QI more fundamentally describes the observable properties that can be probed by a laboratory detector.

\section{Removing the state}\label{sec:nostate}
We have so far tacitly assumed the existence of a predictive state $\op{\rho}$ corresponding to a preparation procedure that is being transformed.  However, in a laboratory any successful (and repeatable) preparation procedure corresponds to a measurement by some detector arrangement.  For example, a polarization state for a laser beam can be prepared by measuring it with a polarizer.  Similarly, the entangled state of a biphoton emitted by a pumped nonlinear crystal via spontaneous parametric down conversion can be prepared by filtering out the remaining pump as a measurement, isolating correlated spatial regions with another measurement, and then performing coincidence filtering as a third measurement.  

Any of these preparation measurements should have a more complete description as a QI in principle.  Thus, we are led to consider the radical possibility that the predictive quantum state also has a more complete description in terms of QIs, at least in principle.  

The key to effectively eliminating the predictive state from the preceding discussion is to observe the role of the identity $\rop{1}$ for obtaining a POM as in Eq.~\eqref{eq:pom}.  This identity indicates an absence of subsequent detections that influence the computed probabilities.  Other detections may be occurring in the laboratory after the detector $\qo{A}$, but none of the computed correlations depend on those detectors.  Hence, they can be entirely omitted in favor of an unbiased final PO: the identity.  

We can perform a similar trick in reverse by conceptually rewinding the preparation procedures to the earliest point that will influence the computed correlation.  Any preparation prior to that first measurement will be irrelevant for later computed probabilities, so we are free to insert the least biased predictive state: the maximally mixed state $\op{\rho} \propto \op{1}$.

By way of example, consider the photodetector in Eq.~\eqref{eq:photo} as a preparation procedure.  After any outcome, the photodetector will update a preparation state to be the vacuum $\pr{0}$.  Hence, any prior state to the operation of the photodetector will be irrelevant for subsequent measurement probabilities.  We can thus declare the input state to the photodetector to be the maximally mixed state and renormalize the total detection probability to eliminate the influence of this choice in a similar way to Eq.~\eqref{eq:condenergy}.  Moreover, the renormalization ratio will eliminate the proportionality constant for the maximally mixed state.

We therefore postulate the following effectively \emph{stateless} reformulation: the measurable joint probability for a sequence of outcomes $(S_1,\cdots,S_k)$ for $k$ detectors will be determined entirely by their QIs according to the ratio
\begin{align}\label{eq:statelessprobs}
  p_{S_1,\ldots,S_k} &= \frac{n_{S_1,\ldots,S_k}}{N} = \frac{\ipr{\rop{1}}{\qo{A}^{(k)}_{S_k}\cdots\qo{A}^{(1)}_{S_1}\op{1}}}{\ipr{\rop{1}}{\qo{A}^{(k)}\cdots\qo{A}^{(1)}\op{1}}},
\end{align}
provided that no neglected measurement prior to $S_1$ or after $S_k$ influences the detections under consideration. Here the $n_{S_1,\ldots,S_k}$ is a positive number that corresponds to the selected detections, while $N > n_{S_1,\ldots,S_k}$ is a normalizing positive number that corresponds to all possible detections.  The non-selective measurements in the denominator and the complete positivity of the QOs that compose the QIs guarantee that this ratio produces properly normalized joint probabilities.  

Importantly, the inclusion of the explicit normalization constant $N$ allows intermediate detectors to have \emph{loss}.  Moreover, the ratio corresponds to the experimental procedure of computing probabilities as ratios of detected events.  In this sense, the stateless expression in Eq.~\eqref{eq:statelessprobs} closely parallels what is being done by an experimenter in a real laboratory situation.

\begin{figure}
  \subfloat[Unnormalized correlation]{
    \begin{tikzpicture}[scale=.4]
      \node[scalar,inner sep=-1mm] (IP) {$c_{\alpha_1,\ldots,\alpha_k}$};
      \node[right=0mm of IP] {=};
      \node[unit] (t) [right=1cm of IP] {};
      \node[costate] (F) [above=of t] {$\rop{1}$} edge [-] (t);
      \node[qo] (Ak) [below=of t] {$\qo{A}^{(k)}[\alpha_k]$} edge [-] (t);
      \node[qo] (Ai) [below=of Ak] {$\cdots$} edge [-] (Ak);
      \node[qo] (A1) [below=of Ai] {$\qo{A}^{(1)}[\alpha_1]$} edge [-] (Ai);
      \node[state] (I) [below=of A1] {$\op{1}$} edge [-] (A1);
    \end{tikzpicture}
    \label{fig:corr}
  }
  \qquad
  \subfloat[Normalization]{
    \begin{tikzpicture}[scale=.4]
      \node[scalar] (IP) {$N$};
      \node[right=0mm of IP] {=};
      \node[unit] (t) [right=1cm of IP] {};
      \node[costate] (F) [above=of t] {$\rop{1}$} edge [-] (t);
      \node[qo] (Ak) [below=of t] {$\qo{A}^{(k)}[1]$} edge [-] (t);
      \node[qo] (Ai) [below=of Ak] {$\cdots$} edge [-] (Ak);
      \node[qo] (A1) [below=of Ai] {$\qo{A}^{(1)}[1]$} edge [-] (Ai);
      \node[state] (I) [below=of A1] {$\op{1}$} edge [-] (A1);
    \end{tikzpicture}
    \label{fig:norm}
  }
  \caption{Stateless reformulation.  (a) The unnormalized correlation $c_{\alpha_1,\ldots,\alpha_k}$ between the contextual values assigned to $k$ detectors $(\qo{A}^{(1)},\cdots,\qo{A}^{(k)})$ in sequence, as in Eq.~\eqref{eq:stateless}.  The quantum instruments for the detectors are entirely sufficient for computing this correlation, provided that no prior or posterior detectors that are not included in the sequence influence the computed correlation. (b) The normalization constant $N$ is computed with the non-selective measurements for the same detector sequence, and produces normalized correlation functions as $\mean{\alpha_1\cdots\alpha_k} = c_{\alpha_1,\ldots,\alpha_k}/N$.  Joint probabilities are special cases when the contextual values are chosen to be indicator functions.}
  \label{fig:stateless}
\end{figure}
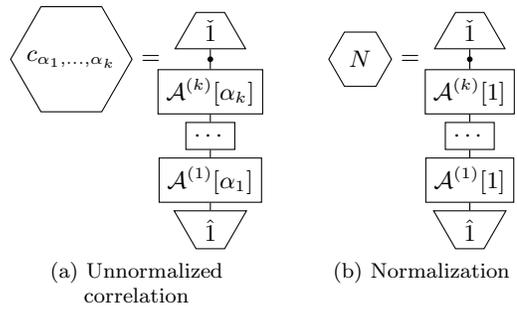

It follows that any measurable correlation between the labels assigned to the $N$ detector outcomes can be computed using the QIs as
\begin{align}\label{eq:stateless}
  \mean{\alpha_1\cdots\alpha_k} &= \frac{c_{\alpha_1,\ldots,\alpha_k}}{N} = \frac{\ipr{\rop{1}}{\qo{A}^{(k)}[\alpha_k]\cdots\qo{A}^{(1)}[\alpha_1]\op{1}}}{\ipr{\rop{1}}{\qo{A}^{(k)}\cdots\qo{A}^{(1)}\op{1}}}.
\end{align}
Here $c_{\alpha_1,\ldots,\alpha_k}$ is an unnormalized correlation between the assigned detector labels as shown in Figure~\ref{fig:corr}, and $N$ is the same normalization factor as computed in Eq.~\eqref{eq:statelessprobs} and shown in Figure~\ref{fig:norm}.

Notably, this reformulation is now \emph{symmetric} in its treatment of the beginning and end points of the computation.  The identity operators $\op{1}$ and $\rop{1}$ remain as place-holders solely to extract the relative magnitudes of the involved operations.  They could in fact be entirely suppressed notationally by defining an average for the QIs directly $\mean{\qo{A}[\alpha]} = \ipr{\rop{1}}{\qo{A}[\alpha]\op{1}}$.  However, the inner product notation will be advantageous in the discussion to follow, so we shall continue to use it.

\section{Retrodictive observables}\label{sec:retroobs}
Now that we have removed the initial state, it is easy to see that a retrodictive quantum observable operator can be defined from a QI in a completely analogous way to a predictive observable, provided that we only consider the \emph{first} detector to be measured.  As shown in Figure~\ref{fig:po}, we first observe that a QO applied to the identity still produces a positive probability operator (PO)
\begin{align}\label{eq:rpo}
  \qo{A}_S\op{1} = \op{A}_S = \int_S\text{d}x\int_Y\text{d}y\, M_{x,y}M^\dagger_{x,y}
\end{align}
that is completely analogous to Eq.~\eqref{eq:po}, but has an inverted ordering of the measurement operators.  We call this a \emph{retrodictive} PO \cite{Pegg1999} for reasons that will become clear in the next section and distinguish it from the predictive PO by the orientation of its hat.

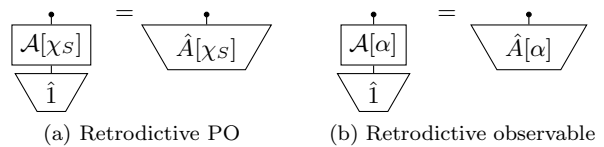
\begin{figure}
  \subfloat[Retrodictive PO]{
  \begin{tikzpicture}[scale=.4]
    \node[unit] (IP) {};
    \node[qo] (A) [below=of IP] {$\qo{A}[\chi_S]$} edge [-] (IP);
    \node[state] (F1) [below=of A] {$\op{1}$} edge [-] (A);
    \node[right=7mm of IP] {=};
    \node[unit] (t) [right=20mm of IP] {};
    \node[state] (F) [below= of t] {$\op{A}[\chi_S]$} edge [-] (t);
  \end{tikzpicture}
  \label{fig:rpo}
  }
  \qquad
  \subfloat[Retrodictive observable]{
  \begin{tikzpicture}[scale=.4]
    \node[unit] (IP) {};
    \node[qo] (A) [below=of IP] {$\qo{A}[\alpha]$} edge [-] (IP);
    \node[state] (F1) [below=of A] {$\op{1}$} edge [-] (A);
    \node[right=7mm of IP] {=};
    \node[unit] (t) [right=20mm of IP] {};
    \node[state] (F) [below= of t] {$\op{A}[\alpha]$} edge [-] (t);
  \end{tikzpicture}
  \label{fig:retrobs}
  }
  \caption{Retrodictive observable operators.  (a) The quantum instrument $\qo{A}$ for the first measured detector generates retrodictive probability operators $\op{A}[\chi_S]$ in a retrodictive POM from indicator functions $\chi_S$ for sets $S$ of detector outcomes, as in Eq.~\eqref{eq:rpo}.  (b) Similarly, weighting the instrument outcomes with real contextual values $\alpha$ produces retrodictive observable operators $\op{A}[\alpha]$ that can be indirectly measured by the first detector, as in Eq.~\eqref{eq:rpom}.}
  \label{fig:retrobservables}
\end{figure}

It follows that a QI for a single measurement will induce a corresponding \emph{retrodictive} POM according to 
\begin{align}\label{eq:rpom}
  \qo{A}[\alpha]\op{1} &= \op{A}[\alpha] = \int_X \alpha(x)\, \textrm{d}\op{A}_x,
\end{align}
as shown in Figure~\ref{fig:retrobs}.  This POM $\op{A}[\alpha]$ is an operator-valued functional of the labels $\alpha(x)$, exactly as the predictive POM in Eq.~\eqref{eq:pom}.  If the QI is \emph{retrodictively complete} then its associated retrodictive POM will form a partition of unity $\op{A}[1] = \op{1}$; however, it is worth noting that predictive completeness does not imply retrodictive completeness.  

As with the predictive POM, a retrodictive POM $\op{A}[\alpha]$ produces a Hermitian observable operator---or \emph{retrodictive observable}---when the chosen contextual values $\alpha$ are real.  However, the predictive and retrodictive observables constructed by the same detector and chosen values will not be equal unless the measurement operators are also Hermitian (and thus positive), so $\rop{A}[\alpha] \neq \op{A}[\alpha]$ in general.

As a quick illustration before we continue, let us revisit our nontrivial photodetector example.  According to its QI in Eq.~\eqref{eq:photo}, the retrodictive POM for the photodetector has the form
\begin{align}\label{eq:photorpo}
  \op{P}[\alpha] &= \qo{P}[\alpha]\op{1} = \pr{0} \left[\sum_{n=0}^{N-1} \alpha_n + \alpha_N \sum_{k=N}^\infty\right].
\end{align}
Unlike the predictive POM in Eq.~\eqref{eq:photopom}, the retrodictive POM consists only of a zero photon projector scaled by a (generally divergent) constant.  Changing the contextual values only changes the value of this constant.

The retrodictive POM of Eq.~\eqref{eq:photorpo} implies that the retrodictive operation of the photodetector is incomplete and strongly biased in spite of the fact that its predictive operation is complete: $\rop{P}[1] = \rop{1}$.  In particular $\op{P}[1] = \aleph_0 \pr{0} \neq \op{1}$, where $\aleph_0 = \Tr{\rop{1}} = \sum_{n=0}^\infty$ formally represents the countable infinity \cite{Robinson1966} of the non-negative integers \footnote{Such an infinite constant can be given rigorous meaning as an element of an expanded number field, such as Robinson's non-standard integers \cite{Robinson1966}.}.  This formally infinite constant seems problematic, but will only appear as a normalization factor in the denominator of expressions like Eq.~\eqref{eq:stateless}.  Therefore, meaningful detection probabilities will still be calculated using the retrodictive POM, such as the unbiased uniform distribution $p_k = \ipr{\rop{1}}{\op{P}[\chi_k]}/\ipr{\rop{1}}{\op{P}[1]} = 1 / \aleph_0$ that is infinitesimal and equal for all $k$, but is also correctly normalized $\sum_{k=0}^\infty p_k = \aleph_0 / \aleph_0 = 1$.  One could also introduce an upper bound to the possible detectable photon numbers as a high-energy cutoff in order to more physically regularize this infinity, if desired, which would introduce loss into the detector description.

\section{Rederiving states}\label{sec:states}
Since we have removed any explicit mention of a quantum state in the reformulation of Eq.~\eqref{eq:stateless}, it is now instructive to examine how states will naturally reappear in calculations.  The symmetric nature of the reformulation allows us to condition the detectable joint probability in different ways.  We find that different choices of conditioning correspond to the appearance of different types of states.

For this purpose it will be sufficient to consider a specific sequence of three detectors, $\qo{A}$, $\qo{B}$, and $\qo{C}$, where we can conceptually understand the middle detector $\mathcal{B}$ as a composition of any number of intermediate detectors.  For simplicity and concreteness we consider each detector to have a discrete number of outcomes, each with a single measurement operator (as will be the typical case in the laboratory)
\begin{subequations}\label{eq:tripledets}
\begin{align}
  \mathcal{A}[\alpha]\op{O} &= \sum_a \alpha_a M_a \op{O} M_a^\dagger, \\
  \mathcal{B}[\beta]\op{O} &= \sum_b \beta_b N_b \op{O} N_b^\dagger, \\
  \mathcal{C}[\gamma]\op{O} &= \sum_c \gamma_c Q_c \op{O} Q_c^\dagger,
\end{align}
\end{subequations}
but this can be easily generalized.  Here $\op{O}$ is an arbitrary operator.

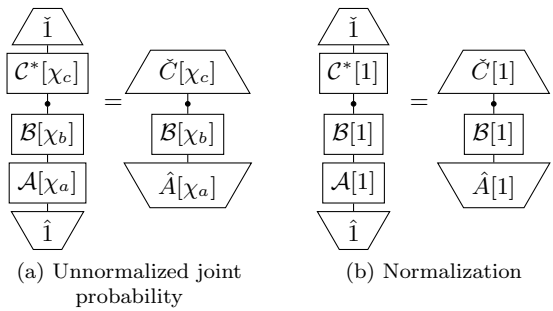
\begin{figure}
  \subfloat[Unnormalized joint probability]{
    \begin{tikzpicture}[scale=.4]
      \node[unit] (t) {};
      \node[qo] (C) [above=of t] {$\qo{C}^*[\chi_c]$} edge [-] (t);
      \node[costate] (F) [above=of C] {$\rop{1}$} edge [-] (C);
      \node[qo] (B) [below=of t] {$\qo{B}[\chi_b]$} edge [-] (t);
      \node[qo] (A) [below=of B] {$\qo{A}[\chi_a]$} edge [-] (B);
      \node[state] (I) [below=of A] {$\op{1}$} edge [-] (A);
      \node[right=6mm of t] {=};
      \node[unit] (t2) [right=18mm of t] {};
      \node[costate] (C2) [above=of t2] {$\rop{C}[\chi_c]$} edge [-] (t2);
      \node[qo] (B2) [below=of t2] {$\qo{B}[\chi_b]$} edge [-] (t2);
      \node[state] (A2) [below=of B2] {$\op{A}[\chi_a]$} edge [-] (B2);
    \end{tikzpicture}
    \label{fig:threecorr}
  }
  \qquad
  \subfloat[Normalization]{
    \begin{tikzpicture}[scale=.4]
      \node[unit] (t) {};
      \node[qo] (C) [above=of t] {$\qo{C}^*[1]$} edge [-] (t);
      \node[costate] (F) [above=of C] {$\rop{1}$} edge [-] (C);
      \node[qo] (B) [below=of t] {$\qo{B}[1]$} edge [-] (t);
      \node[qo] (A) [below=of B] {$\qo{A}[1]$} edge [-] (B);
      \node[state] (I) [below=of A] {$\op{1}$} edge [-] (A);
      \node[right=6mm of t] {=};
      \node[unit] (t2) [right=18mm of t] {};
      \node[costate] (C2) [above=of t2] {$\rop{C}[1]$} edge [-] (t2);
      \node[qo] (B2) [below=of t2] {$\qo{B}[1]$} edge [-] (t2);
      \node[state] (A2) [below=of B2] {$\op{A}[1]$} edge [-] (B2);
    \end{tikzpicture}
    \label{fig:threenorm}
  }
  \caption{Three measurement example.  (a) The unnormalized joint probability $n_{a,b,c}$ as in Eqs.~\eqref{eq:triple} and \eqref{eq:triplesimple} between the contextual values assigned to the sequence of three detectors $(\qo{A},\qo{B},\qo{C})$ defined in Eqs.~\eqref{eq:tripledets}.  The correlation is shown both with quantum instruments and with the associated retrodictive and predictive probability observables for the first and last measurements, respectively. (b) The normalization constant $N$ for the same detector sequence, producing normalized joint probabilities as $p_{a,b,c} = n_{a,b,c}/N$.}
  \label{fig:threeprob}
\end{figure}

The joint probability for obtaining the sequence of outcomes $(a,b,c)$ on these three detectors can then be written according to Eq.~\eqref{eq:statelessprobs} as
\begin{align}\label{eq:triple}
  p_{a,b,c} &= \frac{n_{a,b,c}}{N} = \frac{\ipr{\qo{C}^*_c\rop{1}}{\qo{B}_b\qo{A}_a\op{1}}}{\ipr{\qo{C}^*\rop{1}}{\qo{B}\qo{A}\op{1}}}.
\end{align}
These probabilities are illustrated for reference in Figure~\ref{fig:threeprob}.  For now, we do not assume completeness for the detectors, though we will need to introduce those assumptions later.

The technique for recovering a standard state description is to express the first and last measurements of Eq.~\eqref{eq:triple} in terms of their associated PO.  First we observe that the quantity 
\begin{align}\label{eq:cpo}
  \qo{C}^*_c\rop{1} = \rop{C}_c = Q_c^\dagger Q_c
\end{align}
is the predictive PO for the final measurement as defined in Eq.~\eqref{eq:po} that belongs to the POM $\qo{C}^*[\gamma]\rop{1} = \rop{C}[\gamma]$.  Next we observe that the quantity
\begin{align}\label{eq:arpo}
  \qo{A}_a\op{1} = \op{A}_a = M_a M_a^\dagger
\end{align}
is the retrodictive PO for the first measurement as defined in Eq.~\eqref{eq:rpo} that belongs to the retrodictive POM $\qo{A}[\alpha]\op{1} = \op{A}[\alpha]$.

After these simplifications, Eq.~\eqref{eq:triple} reduces to
\begin{align}\label{eq:triplesimple}
  p_{a,b,c} &= \frac{n_{a,b,c}}{N} = \frac{\ipr{\rop{C}_c}{\qo{B}_b\op{A}_a}}{\ipr{\rop{C}[1]}{\qo{B}\op{A}[1]}},
\end{align}
as shown in Figure~\ref{fig:threeprob}.  The numerator now contains both the retrodictive PO $\op{A}_a$ for the first measurement and the predictive PO $\rop{C}_c$ for the last measurement.  However, the QO $\qo{B}_b$ for the intermediate measurement cannot be replaced by either of its associated POs.  The denominator of \eqref{eq:triplesimple} contains the non-selective bias of the detectors due to loss since we have not assumed completeness of the detectors.  Such a situation was also discussed by Pegg \emph{et al.} \cite{Pegg1999}, albeit without the intermediate measurement.

\subsection{Predictive state}
Let us now consider what happens to the joint probability of Eq.~\eqref{eq:triplesimple} under different strategies of conditioning.  Suppose we wish to condition on a particular outcome $a$ of the first detector as a \emph{preparation} for the remaining two measurements.  This conditioning produces the conditional probabilities
\begin{align}\label{eq:pred}
  p_{b,c|a} &= \frac{p_{a,b,c}}{\sum_{b,c}p_{a,b,c}} = \frac{\ipr{\qo{B}^*_b\rop{C}_c}{\op{A}_a}}{\ipr{\qo{B}^*\rop{C}[1]}{\op{A}_a}}.
\end{align}
The ratio has replaced the previous normalization with a normalization containing only the retrodictive PO $\rop{A}_a$ and the predictive bias $\qo{B}^*\rop{C}[1]$ of the subsequent detectors.  

\begin{figure}
  \begin{tikzpicture}[scale=.4]
    \node[scalar,inner sep=0mm] (p) {$p_{b,c|a}$};
    \node[right=1mm of p] {=};
    \node[unit] (t) [right=14mm of p] {};
    \node[qo] (B) [above=of t] {$\qo{B}^*[\chi_b]$} edge [-] (t);
    \node[costate] (C) [above=of B] {$\rop{C}[\chi_c]$} edge [-] (B);
    \node[state,inner sep=1mm] (A) [below=of t] {$\frac{\op{A}[\chi_a]}{N_a}$} edge [-] (t);
    \node[right=8mm of t] {=};
    \node[unit] (t2) [right=18mm of t] {};
    \node[qo] (B2) [above=of t2] {$\qo{B}^*[\chi_b]$} edge [-] (t2);
    \node[costate] (C2) [above=of B2] {$\rop{C}[\chi_c]$} edge [-] (B2);
    \node[state] (A2) [below=of t2] {$\op{\rho}_a$} edge [-] (t2);
  \end{tikzpicture}
  \caption{Predictive state.  Conditioning on the first measurement and associating the updated normalization constant $N_a = \ipr{\rop{1}}{\op{A}[\chi_a]}$ with the retrodictive PO for the first measurement produces the standard predictive quantum state as in Eq.~\eqref{eq:predstate}, provided that the the detectors $\qo{B}$ and $\qo{C}$ are predictively complete.}
  \label{fig:predstate}
\end{figure}
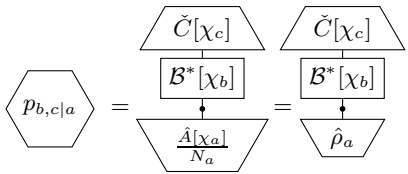

If we also assume that the subsequent detectors are predictively complete, then $\rop{C}[1]=\rop{1}=\rop{B}[1]$ and the denominator simplifies to contain only information about the first detector.  The normalization can then be combined with the PO itself to produce a \emph{predictive state}
\begin{align}\label{eq:predstate}
  \op{\rho}_a &= \frac{\op{A}_a}{\ipr{\rop{1}}{\op{A}_a}}
\end{align}
corresponding entirely to the preparation of the first measurement, as shown in Figure~\ref{fig:predstate}.  

We can therefore understand a predictive state as a renormalized retrodictive PO that corresponds to the \emph{detector} performing the preparation measurement.  It is a proper positive density $\ipr{\rop{1}}{\op{\rho}_a} = 1$ that can be used to predict information about the subsequent measurements, exactly as we used in the first part of this paper.  In terms of this state the conditional probabilities in Eq.~\eqref{eq:pred} have the familiar form
\begin{align}\label{eq:pred2}
  p_{b,c|a} &= \ipr{\qo{B}^*_b\rop{C}_c}{\op{\rho}_a},
\end{align}
where $\qo{B}^*_b\rop{C}_c$ is a joint predictive PO for the final two measurements.  

For the photodetector example, consider conditioning on any particular detector outcome $n$ as a preparation.  Any (possibly infinite) scaling constant from Eq.~\eqref{eq:photorpo} will cancel in Eq.~\eqref{eq:predstate} to yield the predictive state
\begin{align}\label{eq:predphoto}
  \op{\rho}_n &= \frac{\op{P}_n}{\ipr{\rop{1}}{\op{P}_n}} = \pr{0}.
\end{align}
This normalized projector is the proper preparation induced by the photodetector for any outcome, since it always absorbs all available photons.  Thus, the state in Eq.~\eqref{eq:predphoto} predicts that all future photon number measurements will indicate $0$ photons.

\subsection{Retrodictive state}
Alternatively, suppose that we condition on a particular outcome $c$ of the final detector as a \emph{post-selection} for the preceding two measurements.  This conditioning procedure produces the conditional probabilities
\begin{align}\label{eq:retr}
  p_{a,b|c} &= \frac{p_{a,b,c}}{\sum_{a,b}p_{a,b,c}} = \frac{\ipr{\rop{C}_c}{\qo{B}_b\op{A}_a}}{\ipr{\rop{C}_c}{\qo{B}\op{A}[1]}}.
\end{align}
Analogously to Eq.~\eqref{eq:pred} the ratio has replaced the previous normalization with a new normalization that contains only the predictive PO $\rop{C}_c$ and the retrodictive bias $\qo{B}\op{A}[1]$ of the previous detectors.  

\begin{figure}
  \begin{tikzpicture}[scale=.4]
    \node[scalar,inner sep=0mm] (p) {$p_{a,b|c}$};
    \node[right=1mm of p] {=};
    \node[unit] (t) [right=14mm of p] {};
    \node[costate,inner sep=1mm] (C) [above=of t] {$\frac{\rop{C}[\chi_c]}{N_c}$} edge [-] (t);
    \node[qo] (B) [below=of t] {$\qo{B}[\chi_b]$} edge [-] (t);
    \node[state] (A) [below=of B] {$\op{A}[\chi_a]$} edge [-] (B);
    \node[right=8mm of t] {=};
    \node[unit] (t2) [right=18mm of t] {};
    \node[costate] (C2) [above=of t2] {$\rop{\rho}_c$} edge [-] (t2);
    \node[qo] (B2) [below=of t2] {$\qo{B}[\chi_b]$} edge [-] (t2);
    \node[state] (A2) [below=of B2] {$\op{A}[\chi_a]$} edge [-] (B2);
  \end{tikzpicture}
  \caption{Retrodictive state.  Conditioning on the final measurement and associating the updated normalization constant $N_c = \ipr{\rop{C}[\chi_c]}{\op{1}}$ with the PO for the final measurement produces the retrodictive quantum state as in Eq.~\eqref{eq:retrstate}, provided that the detectors $\qo{A}$ and $\qo{B}$ are retrodictively complete.}
  \label{fig:retrstate}
\end{figure}
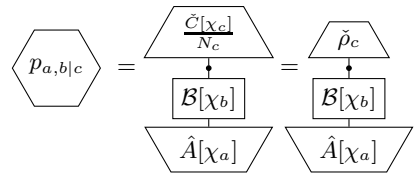

If we assume analogously to the predictive case that the preceding detectors are retrodictively complete, then $\op{A}[1] = \op{1} = \op{B}[1]$ and the denominator simplifies to contain only information about the final detector.  However, we have already seen with the photodetector example in Eq.~\eqref{eq:photorpo} that this assumption may not hold in general, just as the POM completeness assumption used to derive Eq.~\eqref{eq:predstate} may not hold in general.  In the special case with no preceding bias, the new normalization can be combined with the PO itself in an analogous way to Eq.~\eqref{eq:predstate} to produce a \emph{retrodictive state}
\begin{align}\label{eq:retrstate}
  \rop{\rho}_c &= \frac{\rop{C}_c}{\ipr{\rop{C}_c}{\op{1}}}
\end{align}
corresponding entirely to the post-selection of the last measurement, as shown in Figure~\ref{fig:retrstate}.  This definition of a retrodictive state in terms of a predictive PO matches that given by Pegg \emph{et al.} and others \cite{Pegg1999,Barnett2000a,Pegg2002a,Pegg2002b,Chefles2003,Hofmann2003,Pregnell2004,Vaidman2007,Pegg2008,Scroggie2008,Amri2011,Leifer2011}.  The inverted hat reminds us that it gives provides information about prior measurements, rather than predicting future measurements like the predictive state.

We can therefore understand a retrodictive state as a renormalized predictive PO corresponding to the \emph{detector} that performs the post-selection measurement.  The retrodictive state in Eq.~\eqref{eq:retrstate} is a proper positive density $\ipr{\rop{\rho}_c}{\op{1}} = 1$, so satisfies all the standard criteria for a quantum state.  In terms of this state the conditional probabilities in Eq.~\eqref{eq:retr} have the familiar form
\begin{align}\label{eq:retr2}
  p_{a,b|c} &= \ipr{\rop{\rho}_c}{\qo{B}_b\op{A}_a}
\end{align}
analogous to Eq.~\eqref{eq:pred2}, where $\qo{B}_b\op{A}_a$ is a joint retrodictive PO for the preceding two measurements.  

A retrodictive state is used to retroactively infer information about the preceding measurements given known information about the final measurement.  Amri \emph{et al.} \cite{Amri2011} have strongly argued for the interpretation of this state as a \emph{detector} quantity, and have even shown that computing various properties (such as the entanglement) of a retrodictive state will characterize the detection process itself and not any physical system that is being measured.  This observation is thought-provoking in light of the analogous definition in Eq.~\eqref{eq:predstate} for the predictive state.

For the photodetector example, consider conditioning on a particular detector outcome $n\neq N$ as a post-selection.  By using an indicator function with $\alpha_n = 1$ and $\alpha_{n'}=0$ for $n\neq n'$ in the POM of Eq.~\eqref{eq:photopom}, and using the definition in Eq.~\eqref{eq:retrstate} we find the retrodictive state 
\begin{align}\label{eq:photoretrn}
  \rop{\rho}_n &= \frac{\rop{P}_n}{\ipr{\rop{P}_n}{\op{1}}} = \pr{n}.
\end{align}
This normalized projector is the proper post-selection state implied by the photodetector for any definite outcome $n$, since we then know that exactly $n$ photons were absorbed.  

In contrast, for the saturated outcome $N$ we use an indicator function with $\alpha_N=1$ and $\alpha_n=0$ for $n\neq N$ to find the retrodictive state
\begin{align}\label{eq:photoretrN}
  \rop{\rho}_N &= \frac{\rop{P}_N}{\ipr{\rop{P}_N}{\op{1}}} = \left[\frac{1}{\sum_{k=N}^\infty 1}\right] \sum_{k=N}^\infty \pr{k},
\end{align}
which has a formally infinite constant that correctly normalizes a projector onto any photon number of at least $N$.  This retrodictive state is a uniform distribution over the unknown photon numbers that could have been absorbed to produce the $N$\textsuperscript{th} detector outcome, which is the best information that one can infer from the stated operation of the detector given by its QI in Eq.~\eqref{eq:photo}.

\subsection{Time evolution}
Before continuing, we make a brief detour to consider the time evolution of the predictive and retrodictive quantities that have emerged.  To do this, we consider the intermediate measurement $\qo{B}$ to be a simple unitary time-evolution QI $\qo{U}_t$ with a single outcome, exactly as defined in Eq.~\eqref{eq:schrod}.  This replacement effectively reduces our three-measurement sequence to a standard prepare-and-measure scenario, with detector $\qo{A}$ performing the preparation and detector $\qo{C}$ performing the measurement.

We can then rewrite the predictive and retrodictive probabilities from Eqs.~\eqref{eq:pred2} and \eqref{eq:retr2} in the forms
\begin{align}
  \label{eq:predtime}
  p_{c|a} &= \ipr{\rop{C}_c}{\qo{U}_t\op{\rho}_a} = \ipr{\qo{U}^*_t\rop{C}_c}{\op{\rho}_a}, \\
  \label{eq:retrtime}
  p_{a|c} &= \ipr{\rop{\rho}_c}{\qo{U}_t\op{A}_a} = \ipr{\qo{U}^*_t\rop{\rho}_c}{\op{A}_a}.
\end{align}
In both cases, the Schr\"{o}dinger evolution $\qo{U}_t$ can be interpreted as propagating detection information \emph{forward} in time by an interval $t$ from the first measurement to the final measurement, while the Heisenberg evolution $\qo{U}^*_t$ can be interpreted as propagating detection information \emph{backward} in time by an interval $t$ from the final measurement to the first measurement.  The propagation in either temporal direction refers to probabilistic inference, not to a physically propagating object.  The split between the two halves of the inner product in these equations thus indicates a conceptual split between a past and a future with respect to a particular intermediate reference time.

In the predictive case of Eq.~\eqref{eq:predtime} the time-dependent predictive state $\op{\rho}_a(t) = \qo{U}_t\op{\rho}_a$ propagates \emph{forward} in time in the Schr\"{o}dinger picture.  The forward propagating nature of the predictive state is indicated by its upward facing hat.  Similarly, the time-dependent predictive POM $\rop{C}_c(t) = \qo{U}^*_t\rop{C}_c$ (and hence all predictive observables) propagates \emph{backward} in time in the Heisenberg picture.  The inverted evolution of the predictive POM is indicated by the inverted hat.

For the retrodictive case of Eq.~\eqref{eq:retrtime}, on the other hand, the situation is reversed.  The time-dependent retrodictive state $\rop{\rho}_c(t) = \qo{U}^*_t\rop{\rho}_c$ propagates \emph{backward} in time in the Heisenberg picture, while the time-dependent retrodictive POM $\op{A}_a(t) = \qo{U}_t\op{A}_a$ (and hence all retrodictive observables) propagates \emph{forward} in time in the Schr\"{o}dinger picture.  Again, the inverted hat notation indicates which quantities have inverted time evolution.

This conceptual clarification of the Schr\"{o}dinger and Heisenberg pictures of time evolution expands considerably upon the conventional wisdom.  In particular, we now see explicitly that the Heisenberg picture of a predictive time-dependent observable $\rop{C}_c(t)$ implies that the observable corresponds to a \emph{final} measurement in an implicit sequence of two measurements separated by a time \emph{interval} $t$.  Furthermore, this observable is inferentially propagating detection information from a final measurement \emph{backward} in time to be compared with a specific preparation.  For more elaborate detector arrangements this clean separation between a preparation state and a subsequently measured observable will break down, as we shall now emphasize.

\subsection{Interdictive state}
Let us return to our three-measurement sequence and suppose that we condition on a particular \emph{intermediate} detector outcome $b$.  This choice of conditioning produces the conditional probabilities
\begin{align}\label{eq:inter}
  p_{a,c|b} &= \frac{p_{a,b,c}}{\sum_{a,c}p_{a,b,c}} = \frac{\ipr{\rop{C}_c}{\qo{B}_b\op{A}_a}}{\ipr{\rop{C}[1]}{\qo{B}_b\op{A}[1]}}.
\end{align}
Analogously to Eq.~\eqref{eq:pred} and Eq.~\eqref{eq:retr} the ratio has replaced the previous normalization with a new normalization that contains the QO $\qo{B}_b$ for the intermediate measurement as well as the retrodictive and predictive biases $\op{A}[1]$ and $\rop{C}[1]$ of the remaining detectors.  

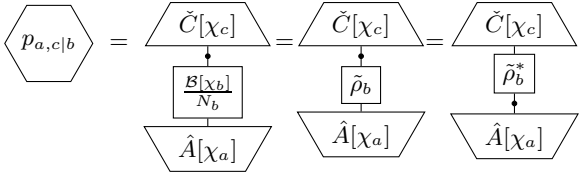
\begin{figure}
  \begin{tikzpicture}[scale=.4]
    \node[scalar,inner sep=0mm] (p) {$p_{a,c|b}$};
    \node[right=1mm of p] {=};
    \node[costate] (C) [right=1cm of p.30] {$\rop{C}[\chi_c]$};
    \node[unit] (t) [below=of C] {} edge [-] (C);
    \node[qo] (B) [below=of t] {$\frac{\qo{B}[\chi_b]}{N_b}$} edge [-] (t);
    \node[state] (A) [below=of B] {$\op{A}[\chi_a]$} edge [-] (B);
    \node[right=23mm of p] {=};
    \node[costate] (C2) [right=7mm of C] {$\rop{C}[\chi_c]$};
    \node[unit] (t2) [below=of C2] {} edge [-] (C2);
    \node[qo] (B2) [below=of t2] {$\tilde{\rho}_b$} edge [-] (t2);
    \node[state] (A2) [below=of B2] {$\op{A}[\chi_a]$} edge [-] (B2);
    \node[right=43mm of p] {=};
    \node[costate] (C3) [right=7mm of C2] {$\rop{C}[\chi_c]$};
    \node[qo] (B3) [below=of C3] {$\tilde{\rho}^*_b$} edge [-] (C3);
    \node[unit] (t3) [below=of B3] {} edge [-] (B3);
    \node[state] (A3) [below=of t3] {$\op{A}[\chi_a]$} edge [-] (t3);
  \end{tikzpicture}
  \caption{Interdictive state.  Conditioning on the intermediate measurement and associating the updated normalization constant $N_b = \ipr{\rop{B}[\chi_b]}{\op{1}} = \ipr{\rop{1}}{\op{B}[\chi_b]}$ with the QO for the intermediate detector produces an interdictive quantum state as in Eq.~\eqref{eq:interstate}, provided that the detectors $\qo{A}$ and $\qo{C}$ are appropriately complete. Unlike the predictive and retrodictive states, the interdictive state is an \emph{operation}.}
  \label{fig:interstate}
\end{figure}

If we assume analogously to the predictive and retrodictive cases that the remaining detectors are appropriately complete, then $\op{A}[1] = \op{1}$ and $\rop{C}[1]=\rop{1}$ and the denominator simplifies to contain only information about the intermediate detector.  In this special case the new normalization can be combined with the QO itself in an analogous way to Eq.~\eqref{eq:predstate} and Eq.~\eqref{eq:retrstate} to produce an \emph{interdictive state} and its adjoint
\begin{align}\label{eq:interstate}
  \tilde{\rho}_b &= \frac{\qo{B}_b}{\ipr{\rop{1}}{\op{B}_b}}, & 
  \tilde{\rho}^*_b &= \frac{\qo{B}^*_b}{\ipr{\rop{B}_b}{\op{1}}},
\end{align}
that correspond entirely to the intermediate detector, as shown in Figure~\ref{fig:interstate}.  Note that the normalization constant for the interdictive state can be written equivalently with the PO or the retrodictive PO associated with the intermediate detector.  In terms of this state the conditional probabilities in Eq.~\eqref{eq:inter} have the compact form
\begin{align}\label{eq:inter2}
  p_{a,c|b} &= \ipr{\rop{C}_c}{\tilde{\rho}_b\op{A}_a} = \ipr{\tilde{\rho}^*_b\rop{C}_c}{\op{A}_a}.
\end{align}

To our knowledge, this sort of state has not appeared in the literature.  Unlike the predictive and retrodictive state operators that appear at the boundaries of the measurement sequence, an interdictive state is a normalized \emph{operation}.  It cannot be written as either the predictive or the retrodictive state operator associated with the intermediate detector, even though it is defined in an entirely analogous way.  We have denoted it with a tilde in Eq.~\eqref{eq:interstate} to distinguish it from the hatted state operators.  The state operation is related to the corresponding state operators according to
\begin{align}\label{eq:interpredretr}
  \op{\rho}_b &= \tilde{\rho}_b\op{1}, & \rop{\rho}_b &= \tilde{\rho}^*_b\rop{1},
\end{align}
so will generally contain more information than either of the associated state operators.

The interdictive state can be used to infer information about both the preceding and subsequent measurements given known information only about an intermediate measurement.  This sort of inference may be appropriate, as an example, for an eavesdropper who wishes to infer information about the correlations between measurements being made at either end of a quantum communication channel.  In such a case the detector $\qo{A}$ would belong to a sender---say, Alice---and the detector $\qo{C}$ would belong to a receiver---say, Charlie.  The intermediate detector $\qo{B}$ would belong to the eavesdropper---say, Beverly---who wishes to learn something about the sort of detectable information being sent through the channel.  Conditioning on each observed outcome $b$ produces the interdictive states accessible by Beverly that indicate what she can infer about the possible correlations between the measurements made by Alice and Charlie after seeing that particular outcome.

For the photodetector example, consider conditioning on a particular detector outcome $n\neq N$ as an intermediate selection.  By using an indicator function with $\alpha_n = 1$ and $\alpha_{n'}=0$ for $n\neq n'$ in the QI of Eq.~\eqref{eq:photo}, and using the definition in Eq.~\eqref{eq:interstate} we find the interdictive state and its adjoint
\begin{subequations}
\begin{align}
  \tilde{\rho}_n\op{O} &= \frac{\qo{P}_n\op{O}}{\ipr{\rop{1}}{\op{P}_n}} = \ket{0}\bra{n}\op{O}\ket{n}\bra{0}, \\
  \tilde{\rho}_n^*\rop{O} &= \frac{\qo{P}^*_n\rop{O}}{\ipr{\rop{P}_n}{\op{1}}} = \ket{n}\bra{0}\rop{O}\ket{0}\bra{n},
\end{align}
\end{subequations}
shown here acting on arbitrary operators $\op{O}$ and $\rop{O}$.  These normalized operations are the proper intermediary selection implied by the photodetector for any definite outcome $n$, since we then know that exactly $n$ photons have been absorbed to leave a vacuum behind.  Applying these interdictive states to an appropriate identity operator according to Eq.~\eqref{eq:interpredretr} correctly recovers the predictive and retrodictive states in Eqs.~\eqref{eq:predphoto} and \eqref{eq:photoretrn}.

In contrast, for the saturated outcome $N$ we use an indicator function with $\alpha_N=1$ and $\alpha_n=0$ for $n\neq N$ to find the interdictive state and its adjoint
\begin{subequations}
\begin{align}
  \tilde{\rho}_N\op{O} &= \frac{\qo{P}_N\op{O}}{\ipr{\rop{1}}{\op{P}_N}} = \frac{\sum_{k=N}^\infty \ket{0}\bra{k}\op{O}\ket{k}\bra{0}}{\sum_{k=N}^\infty 1}, \\
  \tilde{\rho}_N^*\rop{O} &= \frac{\qo{P}^*_N\rop{O}}{\ipr{\rop{P}_N}{\op{1}}} = \frac{\sum_{k=N}^\infty \ket{k}\bra{0}\rop{O}\ket{0}\bra{k}}{\sum_{k=N}^\infty 1},
\end{align}
\end{subequations}
which have formally infinite constants that correctly normalize the operations for any photon number absorption of at least $N$.  Again, applying this interdictive state to an appropriate identity operator according to Eq.~\eqref{eq:interpredretr} recovers the predictive and retrodictive states in Eqs.~\eqref{eq:predphoto} and \eqref{eq:photoretrN}.

\subsection{Bidirectional state}
Finally, let us consider the more subtle case where we condition on \emph{both} a particular preparation $a$ and a particular post-selection $c$ to provide a pre- and post-selection for the intermediate measurement.  This procedure produces the conditional probabilities
\begin{align}\label{eq:bi}
  p_{b|a,c} &= \frac{p_{a,b,c}}{\sum_b p_{a,b,c}} = \frac{\ipr{\rop{C}_c}{\qo{B}_b\op{A}_a}}{\ipr{\rop{C}_c}{\qo{B}\op{A}_a}}.
\end{align}
As with Eqs.~\eqref{eq:pred}, \eqref{eq:retr}, and \eqref{eq:inter}, the normalization has been replaced by the conditioning.  However, the new normalization now contains not only the retrodictive PO $\op{A}_a$ and the predictive PO $\rop{C}_c$ for the pre- and post-selection, but also the \emph{non-selective measurement} $\qo{B}$ for the intermediate measurement.  In this case, assuming completeness for the detectors will not eliminate these dependences from the denominator.  

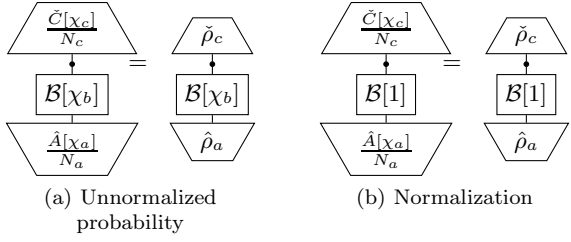
\begin{figure}
  \subfloat[Unnormalized probability]{
    \begin{tikzpicture}[scale=.4]
      \node[unit] (t) {};
      \node[costate,inner sep=1mm] (C) [above=of t] {$\frac{\rop{C}[\chi_c]}{N_c}$} edge [-] (t);
      \node[qo] (B) [below=of t] {$\qo{B}[\chi_b]$} edge [-] (t);
      \node[state,inner sep=1mm] (A) [below=of B] {$\frac{\op{A}[\chi_a]}{N_a}$} edge [-] (B);
      \node[right=6mm of t] {=};
      \node[unit] (t2) [right=18mm of t] {};
      \node[costate] (C2) [above=of t2] {$\rop{\rho}_c$} edge [-] (t2);
      \node[qo] (B2) [below=of t2] {$\qo{B}[\chi_b]$} edge [-] (t2);
      \node[state] (A2) [below=of B2] {$\op{\rho}_a$} edge [-] (B2);
    \end{tikzpicture}
    \label{fig:bicorr}
  }
  \qquad
  \subfloat[Normalization]{
    \begin{tikzpicture}[scale=.4]
      \node[unit] (t) {};
      \node[costate,inner sep=1mm] (C) [above=of t] {$\frac{\rop{C}[\chi_c]}{N_c}$} edge [-] (t);
      \node[qo] (B) [below=of t] {$\qo{B}[1]$} edge [-] (t);
      \node[state,inner sep=1mm] (A) [below=of B] {$\frac{\op{A}[\chi_a]}{N_a}$} edge [-] (B);
      \node[right=6mm of t] {=};
      \node[unit] (t2) [right=18mm of t] {};
      \node[costate] (C2) [above=of t2] {$\rop{\rho}_c$} edge [-] (t2);
      \node[qo] (B2) [below=of t2] {$\qo{B}[1]$} edge [-] (t2);
      \node[state] (A2) [below=of B2] {$\op{\rho}_a$} edge [-] (B2);
    \end{tikzpicture}
    \label{fig:binorm}
  }
  \caption{Bidirectional state.  (a) The unnormalized conditional probability $n_{b|a,c}$ as in Eq.~\eqref{eq:bistate}.  The predictive and retrodictive PO can be independently normalized by convention to produce the pair of states $(\op{\rho}_a,\rop{\rho}_c)$ that encodes the boundary condition information. (b) The normalization constant $N_{a,c}$ that produces normalized conditional probabilities as $p_{b|a,c} = n_{b|a,c}/N_{a,c}$.}
  \label{fig:bistate}
\end{figure}

It is now not so clear how to produce a single state object that fully encapsulates this sort of conditioning.  We can, however, rewrite this probability in terms of both the predictive state $\op{\rho}_a$ and the retrodictive state $\rop{\rho}_c$ by multiplying both numerator and denominator by the appropriate normalization factors used in Eqs.~\eqref{eq:predstate} and \eqref{eq:retrstate}.  We can then treat the pair of states $(\op{\rho}_a,\rop{\rho}_c)$ as a single \emph{bidirectional state}
\begin{align}\label{eq:bistate}
  p_{b|a,c} &= \frac{n_{b|a,c}}{N_{a,c}} = \frac{\ipr{\rop{\rho}_c}{\qo{B}_b\op{\rho}_a}}{\ipr{\rop{\rho}_c}{\qo{B}\op{\rho}_a}}
\end{align}
that specifies the bias at the \emph{boundaries} of the measurement sequence, as shown in Figure~\ref{fig:bistate}.

The choice to normalize each half of the bidirectional state separately is a matter of convention that loses some information: the equivalent form of Eq.~\eqref{eq:bi} using POs retains more detector information, while the QI form in Eq.~\eqref{eq:triple} retains the complete detector information.  In all three forms of the probability, however, the non-selective measurement involving the QI of the intermediate detection is required for proper normalization.  In our opinion, the predictive and retrodictive states are auxiliary objects that can be introduced when the full QIs that describe the pre- and post-selection measurements are not known or needed.  We also note that the probabilities in Eq.~\eqref{eq:bistate} fully generalize the Aharonov-Bergmann-Lebowitz (ABL) rule \cite{Aharonov1964} for pre- and post-selected projective measurements to arbitrary detectors.  

To emphasize the irreducible role of the intermediate measurement, consider the measurable pre- and post-selected average $\ccmean{c}{\beta}{a}$ of a chosen set of detector labels $\beta_b$ for the intermediate measurement $\qo{B}$:
\begin{align}\label{eq:prepostav}
  \ccmean{c}{\beta}{a} &= \sum_b \beta_b\, p_{b|a,c} = \frac{\ipr{\rop{\rho}_c}{\qo{B}[\beta]\op{\rho}_a}}{\ipr{\rop{\rho}_c}{\qo{B}\op{\rho}_a}}, \\
  &= \frac{\sum_b \beta_b \,\Tr{\rop{\rho}_c N_b \op{\rho}_a N_b^\dagger }}{\sum_b \Tr{\rop{\rho}_c N_b \op{\rho}_a N_b^\dagger }}. \nonumber
\end{align}
Neither the predictive $\rop{B}[\beta]$ nor the retrodictive $\op{B}[\beta]$ observable operators associated with the observable \emph{operation} $\qo{B}[\beta]$ can describe this measurable average.  One needs the full QI of the second measurement to construct both the operation $\qo{B}[\beta]$ in the numerator and the non-selective measurement $\qo{B}$ in the denominator.  

For comparison, the stateless form of the average in Eq.~\eqref{eq:prepostav} that uses the QIs as in Eq.~\eqref{eq:triple} is
\begin{align}\label{eq:prepostavmops}
 \ccmean{c}{\beta}{a} &= \frac{\sum_b \beta_b \,\Tr{(Q_c N_b M_a)^\dagger(Q_c N_b M_a)}}{\sum_b \Tr{(Q_c N_b M_a)^\dagger(Q_c N_b M_a)}},
\end{align}
which involves only the Hermitian squares of the composite measurement operator $(Q_c N_b M_a)$ for the measurement sequence.

It is possible, however, to restore either of the observable operators $\rop{B}[\beta]$ or $\op{B}[\beta]$ to Eq.~\eqref{eq:prepostav} in a limited sense by using the identities
\begin{subequations}
\begin{align}
  N_b \op{\rho}_a N_b^\dagger &= \frac{\op{B}_b \op{\rho}_a + \op{\rho}_a \op{B}_b}{2} + \qo{L}[N_b]\op{\rho}_a \\
  N_b^\dagger \rop{\rho}_c N_b &= \frac{\rop{B}_b \rop{\rho}_c + \rop{\rho}_c \rop{B}_b}{2} + \qo{L}[N_b^\dagger]\rop{\rho}_c,
\end{align}
\end{subequations}
where the operation
\begin{align}
  \qo{L}[N_b]\op{O} &= -\frac{1}{2} \left(N_b [N_b^\dagger,\op{O}] - [N_b,\op{O}]N_b^\dagger\right),
\end{align}
involving commutators $[\cdot,\cdot]$ with an arbitrary operator $\op{O}$ is the \emph{Lindblad operation} familiar from studies of decoherence in open quantum systems \cite{Breuer2007,Wiseman2009}.  

The Lindblad operation indicates \emph{disturbance} that the intermediate measurement introduces to the measurement sequence.  We can infer this fact by observing that the non-selective measurement $\qo{B}$ in the denominator of Eq.~\eqref{eq:prepostav} will reduce to the identity operation when the Lindblad terms are neglected due to the assumed completeness relations $\rop{B}[1] = \rop{1}$ and $\op{B}[1] = \op{1}$.  Furthermore, in the symmetric case when $N_b^\dagger = N_b$ (and thus $\rop{B}_b = \op{B}_b$) for all $b$, neglecting the Lindblad terms would make the intermediate measurement completely equivalent to classical Bayesian conditioning, as we showed in \cite{Dressel2012b,Dressel2013}.

This symmetric situation when $\rop{B}[\beta] = \op{B}[\beta] = B[\beta]$ and the Lindblad disturbance can be approximately neglected corresponds to the Aharonov-Albert-Vaidman (AAV) \emph{weak measurement regime} \cite{Aharonov1988,Duck1989}.  In this regime, the intermediate measurement does not appreciably influence the surrounding measurements.  Only in this case can the average in Eq.~\eqref{eq:prepostav} be described entirely by an observable operator $B[\beta]$ and become independent of the QI for the intermediate measurement.  In such a case, the average approximates a \emph{generalized weak value} \cite{Dressel2010,Dressel2011,Dressel2012,Dressel2012b,Dressel2012c,Dressel2012d,Dressel2012e,Dressel2013}
\begin{align}\label{eq:weakvalue}
  \ccmean{c}{\beta}{a} &\approx  \text{Re}\frac{\Tr{\rop{\rho}_c\, B[\beta]\, \op{\rho}_a}}{\Tr{\rop{\rho}_c\, \op{\rho}_a}}.
\end{align}
However, we emphasize that this is an idealized limit that can only be approximated under special conditions.  Generally the disturbance inherent to a quantum measurement cannot be removed, so the full \emph{measurable} average in Eq.~\eqref{eq:prepostav} must be used to properly describe what will be observed in the laboratory.

\section{Reformulation restated}\label{sec:restate}
Returning to our effectively stateless reformulation of Eq.~\eqref{eq:stateless}, we can now consider reintroducing states as a pragmatic convenience.  It may be that an experimenter does not know or care about the origin of boundary bias for a measurement sequence, so does not wish to describe that bias as an appropriate sequence of QIs.  For example, an experimenter may only care that the predictive state corresponding to the output of a single mode laser can be approximated by a paraxial coherent state with a maximally mixed phase and may not wish to describe in more detail how that state may be produced via the interaction between the ignored gain medium and the electromagnetic field \cite{Pegg2005}. 

In such a case, any boundary bias from detectors that are not explicitly included in the self-contained sequence considered in Eq.~\eqref{eq:stateless} can be added by replacing the unbiased identity operators with an appropriate bidirectional state pair $(\op{\rho}_i,\rop{\rho}_f)$:
\begin{align}\label{eq:statelessbiased}
  \mean{\alpha_1\cdots\alpha_k} &= \frac{\ipr{\rop{\rho}_f}{\qo{A}^{(k)}[\alpha_k]\cdots\qo{A}^{(1)}[\alpha_1]\op{\rho}_i}}{\ipr{\rop{\rho}_f}{\qo{A}^{(k)}\cdots\qo{A}^{(1)}\op{\rho}_i}}.
\end{align}
Such a replacement voluntarily discards more detailed information about preparation and post-selection detectors in favor of equivalence classes that are sufficient approximations for computing the measurable correlations.

If there is also intermediate bias from one or more intermediate filtering operations, then one can add any number of interdictive states to Eq.~\eqref{eq:statelessbiased} as needed:
\begin{align}\label{eq:statelessbiasedcomplete}
  \mean{\alpha_1\cdots\alpha_k} &= \frac{\ipr{\rop{\rho}_f}{\qo{A}^{(k)}[\alpha_k]\cdots\tilde{\rho}\cdots\qo{A}^{(1)}[\alpha_1]\op{\rho}_i}}{\ipr{\rop{\rho}_f}{\qo{A}^{(k)}\cdots\tilde{\rho}\cdots\qo{A}^{(1)}\op{\rho}_i}}.
\end{align}
In this case a longer sequence of states $(\op{\rho}_i,\tilde{\rho},\rop{\rho}_f)$ will encode the complete bias that influences the measured detectors.

We emphasize, however, that although the forms of Eqs. \eqref{eq:statelessbiased} and \eqref{eq:statelessbiasedcomplete} may be more practical for laboratory computations, they can be derived from the completely stateless formulation of Eq.~\eqref{eq:stateless} by appropriate conditioning of more detailed physical descriptions using instruments.  In principle, therefore, quantum instruments form a complete foundation for describing any measurable laboratory correlation.  We also emphasize that these expressions are completely general for any sequence of any number of laboratory detectors that may or may not include loss.

\section{Conclusion}\label{sec:conclusion}
In this paper we have discussed the quantum instrument as a foundation for understanding real measurable probabilities and correlations in a laboratory setting.  We showed how the concept of a quantum instrument subsumes the usual concept of a predictive quantum observable and generalizes it to include the transformative effects of the detecting apparatus.  We also showed how one can effectively remove the quantum state from the picture entirely, at least in principle, in favor of quantum instruments that directly correspond to laboratory equipment.  

The resulting stateless reformulation in terms of instruments has the benefit of treating the start and end points of computations symmetrically.  This symmetry permitted us to show how different conditioning strategies for the detected information produce different ideas of a quantum state, as well as different ideas of a quantum observable.  Conditioning on a preparation event produces a predictive state $\op{\rho}$, while conditioning on a posterior post-selection event produces a retrodictive state $\rop{\rho}$.  In two-measurement sequences, standard predictive observables pair naturally with predictive states, while newly appearing retrodictive observables pair naturally with retrodictive states.  In both cases, the states appear as a consequence of conditioning on the outcome of a detector, so we argue that these states should be understood as describing information pertaining to the detectors themselves.  This point of view contrasts sharply with the conventional tendency to interpret a predictive state as intrinsic information about some object that is independent of the detector arrangement.

We explored the time evolution of the emergent predictive and retrodictive quantities by considering time evolution as a special case of a quantum instrument.  We found that Heisenberg evolution is obeyed by both predictive observables and retrodictive states, and corresponds to backward propagation by an interval $t$ of the measurement information contained in posterior measurement events.  Similarly, Schr\"{o}dinger evolution is obeyed by both retrodictive observables and predictive states, and corresponds to forward propagation of the measurement information contained in prior measurement events.  In neither case does our stateless picture imply that these evolving quantities correspond to any physical object propagating between the detectors either forward or backward in time; only detector information is being evolved to give us \emph{inferences} regarding other possible detector events.

In addition to deriving predictive and retrodictive pictures for a two-measurement sequence, our stateless reformulation also allowed us to consider the more subtle situation of a three-measurement sequence.  In this case, there are two new types of conditioning that can occur, which produce new types of state.  First, conditioning on an intermediate event produces an \emph{interdictive} state $\tilde{\rho}$ and its adjoint $\tilde{\rho}^*$ that have the form of normalized \emph{operations}.  This type of state could be useful for the inferences made by an eavesdropper.  Second, conditioning on both a preparation event and a post-selection event produces a \emph{bidirectional state}, which can be represented as a pair $(\op{\rho},\rop{\rho})$ of predictive and retrodictive states corresponding to the induced detection bias at the boundaries of the measurement sequence.  

The resulting conditional probabilities generated from such a bidirectional state contain and generalize the Aharonov-Bergmann-Lebowitz (ABL) rule \cite{Aharonov1964} for projective pre- and post-selected measurement probabilities.  Furthermore, the inclusion of bidirectional boundary bias prevents the emergence of a standard observable operator for the intermediate measurement; nevertheless, the quantum instrument for the intermediate measurement still permits us to construct a measurable pre- and post-selected average.  This measurable average contains and generalizes the Aharonov-Albert-Vaidman (AAV) weak value \cite{Aharonov1988,Duck1989}.  Our reformulation with quantum instruments and bidirectional states thus provides a full generalization and clarification of the ``two-state vector formalism'' of Aharonov \emph{et al.} \cite{Aharonov2008,Aharonov2009,Aharonov2010}.  This generalization enables calculations with mixed ``pre- and post-selection states,'' identifies the post-selection state as a retrodictive state that indicates posterior bias (rather than a retro-causal object), and permits the inclusion of multiple intermediate measurements with proper disturbance and loss.  Importantly, our generalization always explicitly describes measurable laboratory situations and not counter-factual hypotheticals.

For future work that builds on what we have shown here, we point out that our derived forms for the interdictive and bidirectional states do not preclude the possibility of other equivalent descriptions for the relevant bias information.  The work of Leifer, Spekkens, and Coecke using conditional density operators \cite{Leifer2011,Coecke2012} may provide clues for additional investigation along these lines.  As an example, one could consider constructing a minimal representation for a bidirectional state as an equivalence class of all the bidirectional state pairs that produce identical probabilities for any intermediate measurement.  Such a minimal representation for a bidirectional state would closely connect to work by Crutchfield \emph{et al.} in the characterization of classical stochastic processes \cite{Crutchfield2009,Ellison2009,Ellison2011}, who conclude that such a representation will most efficiently encode the accessible information that one can infer solely from observing realizations of a process.  However, they also notably conclude that this encoded information may still be insufficient for modeling an underlying mechanism that generates the stochastic sequence---even if there is one. Extrapolating this observation to the quantum realm has interesting implications for the continued efforts to postulate an ontological mechanism that generates the apparent randomness intrinsic to the measurement process.

\begin{acknowledgments}
  We thank Curtis Broadbent for suggesting the term ``interdictive state.''  We acknowledge support from the National Science Foundation under Grant No. DMR-0844899, and the US Army Research Office under grant Grant No. W911NF-09-0-01417.
\end{acknowledgments}


\begin{thebibliography}{72}%
\makeatletter
\providecommand \@ifxundefined [1]{%
 \@ifx{#1\undefined}
}%
\providecommand \@ifnum [1]{%
 \ifnum #1\expandafter \@firstoftwo
 \else \expandafter \@secondoftwo
 \fi
}%
\providecommand \@ifx [1]{%
 \ifx #1\expandafter \@firstoftwo
 \else \expandafter \@secondoftwo
 \fi
}%
\providecommand \natexlab [1]{#1}%
\providecommand \enquote  [1]{``#1''}%
\providecommand \bibnamefont  [1]{#1}%
\providecommand \bibfnamefont [1]{#1}%
\providecommand \citenamefont [1]{#1}%
\providecommand \href@noop [0]{\@secondoftwo}%
\providecommand \href [0]{\begingroup \@sanitize@url \@href}%
\providecommand \@href[1]{\@@startlink{#1}\@@href}%
\providecommand \@@href[1]{\endgroup#1\@@endlink}%
\providecommand \@sanitize@url [0]{\catcode `\\12\catcode `\$12\catcode
  `\&12\catcode `\#12\catcode `\^12\catcode `\_12\catcode `\%12\relax}%
\providecommand \@@startlink[1]{}%
\providecommand \@@endlink[0]{}%
\providecommand \url  [0]{\begingroup\@sanitize@url \@url }%
\providecommand \@url [1]{\endgroup\@href {#1}{\urlprefix }}%
\providecommand \urlprefix  [0]{URL }%
\providecommand \Eprint [0]{\href }%
\providecommand \doibase [0]{http://dx.doi.org/}%
\providecommand \selectlanguage [0]{\@gobble}%
\providecommand \bibinfo  [0]{\@secondoftwo}%
\providecommand \bibfield  [0]{\@secondoftwo}%
\providecommand \translation [1]{[#1]}%
\providecommand \BibitemOpen [0]{}%
\providecommand \bibitemStop [0]{}%
\providecommand \bibitemNoStop [0]{.\EOS\space}%
\providecommand \EOS [0]{\spacefactor3000\relax}%
\providecommand \BibitemShut  [1]{\csname bibitem#1\endcsname}%
\let\auto@bib@innerbib\@empty
\bibitem [{\citenamefont {Dirac}(1930)}]{Dirac1930}%
  \BibitemOpen
  \bibfield  {author} {\bibinfo {author} {\bibfnamefont {P.~A.~M.}\
  \bibnamefont {Dirac}},\ }\href@noop {} {\emph {\bibinfo {title} {{Principles
  of Quantum Mechanics}}}}\ (\bibinfo  {publisher} {Oxford University Press,
  Oxford},\ \bibinfo {year} {1930})\BibitemShut {NoStop}%
\bibitem [{\citenamefont {von Neumann}(1932)}]{VonNeumann1932}%
  \BibitemOpen
  \bibfield  {author} {\bibinfo {author} {\bibfnamefont {J.}~\bibnamefont {von
  Neumann}},\ }\href@noop {} {\emph {\bibinfo {title} {Mathematische Grundlagen
  der Quantenmechanik}}}\ (\bibinfo  {publisher} {Berlin: Springer},\ \bibinfo
  {year} {1932})\BibitemShut {NoStop}%
\bibitem [{\citenamefont {Alicki}\ and\ \citenamefont
  {Fannes}(2001)}]{Alicki2001}%
  \BibitemOpen
  \bibfield  {author} {\bibinfo {author} {\bibfnamefont {R.}~\bibnamefont
  {Alicki}}\ and\ \bibinfo {author} {\bibfnamefont {M.}~\bibnamefont
  {Fannes}},\ }\href@noop {} {\emph {\bibinfo {title} {Quantum dynamical
  systems}}}\ (\bibinfo  {publisher} {Oxford University Press, Oxford},\
  \bibinfo {year} {2001})\BibitemShut {NoStop}%
\bibitem [{\citenamefont {Wiseman}\ and\ \citenamefont
  {Milburn}(2009)}]{Wiseman2009}%
  \BibitemOpen
  \bibfield  {author} {\bibinfo {author} {\bibfnamefont {H.~M.}\ \bibnamefont
  {Wiseman}}\ and\ \bibinfo {author} {\bibfnamefont {G.}~\bibnamefont
  {Milburn}},\ }\href@noop {} {\emph {\bibinfo {title} {Quantum Measurement and
  Control}}}\ (\bibinfo  {publisher} {Cambridge University Press, Cambridge},\
  \bibinfo {year} {2009})\BibitemShut {NoStop}%
\bibitem [{\citenamefont {Jaynes}(2003)}]{Jaynes2003}%
  \BibitemOpen
  \bibfield  {author} {\bibinfo {author} {\bibfnamefont {E.~T.}\ \bibnamefont
  {Jaynes}},\ }\href@noop {} {\emph {\bibinfo {title} {{Probability Theory: The
  Logic of Science}}}}\ (\bibinfo  {publisher} {Cambridge University Press,
  Cambridge},\ \bibinfo {year} {2003})\BibitemShut {NoStop}%
\bibitem [{\citenamefont {Jauch}(1968)}]{Jauch1968}%
  \BibitemOpen
  \bibfield  {author} {\bibinfo {author} {\bibfnamefont {J.~M.}\ \bibnamefont
  {Jauch}},\ }\href@noop {} {\emph {\bibinfo {title} {Foundations of quantum
  mechanics}}}\ (\bibinfo  {publisher} {Addison-Wesley Pub. Co., Reading},\
  \bibinfo {year} {1968})\BibitemShut {NoStop}%
\bibitem [{\citenamefont {Dressel}\ and\ \citenamefont
  {Jordan}(2012{\natexlab{a}})}]{Dressel2012b}%
  \BibitemOpen
  \bibfield  {author} {\bibinfo {author} {\bibfnamefont {J.}~\bibnamefont
  {Dressel}}\ and\ \bibinfo {author} {\bibfnamefont {A.~N.}\ \bibnamefont
  {Jordan}},\ }\href@noop {} {\bibfield  {journal} {\bibinfo  {journal} {Phys.
  Rev. A}\ }\textbf {\bibinfo {volume} {85}},\ \bibinfo {pages} {022123}
  (\bibinfo {year} {2012}{\natexlab{a}})}\BibitemShut {NoStop}%
\bibitem [{\citenamefont {Dressel}(2013)}]{Dressel2013}%
  \BibitemOpen
  \bibfield  {author} {\bibinfo {author} {\bibfnamefont {J.}~\bibnamefont
  {Dressel}},\ }\emph {\bibinfo {title} {{Indirect Observable Measurement: an
  Algebraic Approach}}},\ \href@noop {} {Ph.D. thesis},\ \bibinfo  {school}
  {{University of Rochester}} (\bibinfo {year} {2013})\BibitemShut {NoStop}%
\bibitem [{\citenamefont {Watanabe}(1955)}]{Watanabe1955}%
  \BibitemOpen
  \bibfield  {author} {\bibinfo {author} {\bibfnamefont {S.}~\bibnamefont
  {Watanabe}},\ }\href@noop {} {\bibfield  {journal} {\bibinfo  {journal} {Rev.
  Mod. Phys.}\ }\textbf {\bibinfo {volume} {27}},\ \bibinfo {pages} {179}
  (\bibinfo {year} {1955})}\BibitemShut {NoStop}%
\bibitem [{\citenamefont {Aharonov}\ \emph {et~al.}(1964)\citenamefont
  {Aharonov}, \citenamefont {Bergmann},\ and\ \citenamefont
  {Lebowitz}}]{Aharonov1964}%
  \BibitemOpen
  \bibfield  {author} {\bibinfo {author} {\bibfnamefont {Y.}~\bibnamefont
  {Aharonov}}, \bibinfo {author} {\bibfnamefont {P.~G.}\ \bibnamefont
  {Bergmann}}, \ and\ \bibinfo {author} {\bibfnamefont {J.~L.}\ \bibnamefont
  {Lebowitz}},\ }\href@noop {} {\bibfield  {journal} {\bibinfo  {journal}
  {Phys. Rev}\ }\textbf {\bibinfo {volume} {134}},\ \bibinfo {pages} {B1410}
  (\bibinfo {year} {1964})}\BibitemShut {NoStop}%
\bibitem [{\citenamefont {Pegg}\ and\ \citenamefont
  {Barnett}(1999)}]{Pegg1999}%
  \BibitemOpen
  \bibfield  {author} {\bibinfo {author} {\bibfnamefont {D.~T.}\ \bibnamefont
  {Pegg}}\ and\ \bibinfo {author} {\bibfnamefont {S.~M.}\ \bibnamefont
  {Barnett}},\ }\href@noop {} {\bibfield  {journal} {\bibinfo  {journal} {J.
  Opt. B: Quantum Semiclass. Opt.}\ }\textbf {\bibinfo {volume} {1}},\ \bibinfo
  {pages} {442} (\bibinfo {year} {1999})}\BibitemShut {NoStop}%
\bibitem [{\citenamefont {Barnett}\ \emph
  {et~al.}(2000{\natexlab{a}})\citenamefont {Barnett}, \citenamefont {Pegg},\
  and\ \citenamefont {Jeffers}}]{Barnett2000a}%
  \BibitemOpen
  \bibfield  {author} {\bibinfo {author} {\bibfnamefont {S.~M.}\ \bibnamefont
  {Barnett}}, \bibinfo {author} {\bibfnamefont {D.~T.}\ \bibnamefont {Pegg}}, \
  and\ \bibinfo {author} {\bibfnamefont {J.}~\bibnamefont {Jeffers}},\
  }\href@noop {} {\bibfield  {journal} {\bibinfo  {journal} {J. Mod. Opt.}\
  }\textbf {\bibinfo {volume} {47}},\ \bibinfo {pages} {1779} (\bibinfo {year}
  {2000}{\natexlab{a}})}\BibitemShut {NoStop}%
\bibitem [{\citenamefont {Pegg}\ \emph
  {et~al.}(2002{\natexlab{a}})\citenamefont {Pegg}, \citenamefont {Barnett},\
  and\ \citenamefont {Jeffers}}]{Pegg2002a}%
  \BibitemOpen
  \bibfield  {author} {\bibinfo {author} {\bibfnamefont {D.}~\bibnamefont
  {Pegg}}, \bibinfo {author} {\bibfnamefont {S.}~\bibnamefont {Barnett}}, \
  and\ \bibinfo {author} {\bibfnamefont {J.}~\bibnamefont {Jeffers}},\
  }\href@noop {} {\bibfield  {journal} {\bibinfo  {journal} {J. Mod. Opt.}\
  }\textbf {\bibinfo {volume} {49}},\ \bibinfo {pages} {913} (\bibinfo {year}
  {2002}{\natexlab{a}})}\BibitemShut {NoStop}%
\bibitem [{\citenamefont {Pegg}\ \emph
  {et~al.}(2002{\natexlab{b}})\citenamefont {Pegg}, \citenamefont {Barnett},\
  and\ \citenamefont {Jeffers}}]{Pegg2002b}%
  \BibitemOpen
  \bibfield  {author} {\bibinfo {author} {\bibfnamefont {D.~T.}\ \bibnamefont
  {Pegg}}, \bibinfo {author} {\bibfnamefont {S.~M.}\ \bibnamefont {Barnett}}, \
  and\ \bibinfo {author} {\bibfnamefont {J.}~\bibnamefont {Jeffers}},\
  }\href@noop {} {\bibfield  {journal} {\bibinfo  {journal} {Phys. Rev. A}\
  }\textbf {\bibinfo {volume} {66}},\ \bibinfo {pages} {022106} (\bibinfo
  {year} {2002}{\natexlab{b}})}\BibitemShut {NoStop}%
\bibitem [{\citenamefont {Chefles}\ and\ \citenamefont
  {Sasaki}(2003)}]{Chefles2003}%
  \BibitemOpen
  \bibfield  {author} {\bibinfo {author} {\bibfnamefont {A.}~\bibnamefont
  {Chefles}}\ and\ \bibinfo {author} {\bibfnamefont {M.}~\bibnamefont
  {Sasaki}},\ }\href@noop {} {\bibfield  {journal} {\bibinfo  {journal} {Phys.
  Rev. A}\ }\textbf {\bibinfo {volume} {67}},\ \bibinfo {pages} {032112}
  (\bibinfo {year} {2003})}\BibitemShut {NoStop}%
\bibitem [{\citenamefont {Hofmann}(2003)}]{Hofmann2003}%
  \BibitemOpen
  \bibfield  {author} {\bibinfo {author} {\bibfnamefont {H.}~\bibnamefont
  {Hofmann}},\ }\href@noop {} {\bibfield  {journal} {\bibinfo  {journal} {Phys.
  Rev. A}\ }\textbf {\bibinfo {volume} {67}},\ \bibinfo {pages} {022106}
  (\bibinfo {year} {2003})}\BibitemShut {NoStop}%
\bibitem [{\citenamefont {Pregnell}\ and\ \citenamefont
  {Pegg}(2004)}]{Pregnell2004}%
  \BibitemOpen
  \bibfield  {author} {\bibinfo {author} {\bibfnamefont {K.~T.}\ \bibnamefont
  {Pregnell}}\ and\ \bibinfo {author} {\bibfnamefont {D.~T.}\ \bibnamefont
  {Pegg}},\ }\href@noop {} {\bibfield  {journal} {\bibinfo  {journal} {J. Mod.
  Opt.}\ }\textbf {\bibinfo {volume} {51}},\ \bibinfo {pages} {1613} (\bibinfo
  {year} {2004})}\BibitemShut {NoStop}%
\bibitem [{\citenamefont {Vaidman}(2007)}]{Vaidman2007}%
  \BibitemOpen
  \bibfield  {author} {\bibinfo {author} {\bibfnamefont {L.}~\bibnamefont
  {Vaidman}},\ }\href@noop {} {\bibfield  {journal} {\bibinfo  {journal} {J.
  Phys. A: Math. Theor.}\ }\textbf {\bibinfo {volume} {40}},\ \bibinfo {pages}
  {3275} (\bibinfo {year} {2007})}\BibitemShut {NoStop}%
\bibitem [{\citenamefont {Pegg}(2008)}]{Pegg2008}%
  \BibitemOpen
  \bibfield  {author} {\bibinfo {author} {\bibfnamefont {D.~T.}\ \bibnamefont
  {Pegg}},\ }\href@noop {} {\bibfield  {journal} {\bibinfo  {journal} {Found.
  Phys.}\ }\textbf {\bibinfo {volume} {38}},\ \bibinfo {pages} {648} (\bibinfo
  {year} {2008})}\BibitemShut {NoStop}%
\bibitem [{\citenamefont {Scroggie}\ and\ \citenamefont
  {Jeffers}(2008)}]{Scroggie2008}%
  \BibitemOpen
  \bibfield  {author} {\bibinfo {author} {\bibfnamefont {A.~J.}\ \bibnamefont
  {Scroggie}}\ and\ \bibinfo {author} {\bibfnamefont {J.}~\bibnamefont
  {Jeffers}},\ }\href@noop {} {\bibfield  {journal} {\bibinfo  {journal} {Int.
  J. Theor. Phys.}\ }\textbf {\bibinfo {volume} {47}},\ \bibinfo {pages} {1809}
  (\bibinfo {year} {2008})}\BibitemShut {NoStop}%
\bibitem [{\citenamefont {Amri}\ \emph {et~al.}(2011)\citenamefont {Amri},
  \citenamefont {Laurat},\ and\ \citenamefont {Fabre}}]{Amri2011}%
  \BibitemOpen
  \bibfield  {author} {\bibinfo {author} {\bibfnamefont {T.}~\bibnamefont
  {Amri}}, \bibinfo {author} {\bibfnamefont {J.}~\bibnamefont {Laurat}}, \ and\
  \bibinfo {author} {\bibfnamefont {C.}~\bibnamefont {Fabre}},\ }\href@noop {}
  {\bibfield  {journal} {\bibinfo  {journal} {Phys. Rev. Lett.}\ }\textbf
  {\bibinfo {volume} {106}},\ \bibinfo {pages} {020502} (\bibinfo {year}
  {2011})}\BibitemShut {NoStop}%
\bibitem [{\citenamefont {Barnett}\ \emph
  {et~al.}(2000{\natexlab{b}})\citenamefont {Barnett}, \citenamefont {Pegg},
  \citenamefont {Jeffers},\ and\ \citenamefont {Jedrkiewicz}}]{Barnett2000b}%
  \BibitemOpen
  \bibfield  {author} {\bibinfo {author} {\bibfnamefont {S.~M.}\ \bibnamefont
  {Barnett}}, \bibinfo {author} {\bibfnamefont {D.~T.}\ \bibnamefont {Pegg}},
  \bibinfo {author} {\bibfnamefont {J.}~\bibnamefont {Jeffers}}, \ and\
  \bibinfo {author} {\bibfnamefont {O.}~\bibnamefont {Jedrkiewicz}},\
  }\href@noop {} {\bibfield  {journal} {\bibinfo  {journal} {J. Phys. B: At.
  Mol. Opt. Phys.}\ }\textbf {\bibinfo {volume} {33}},\ \bibinfo {pages} {3047}
  (\bibinfo {year} {2000}{\natexlab{b}})}\BibitemShut {NoStop}%
\bibitem [{\citenamefont {Barnett}\ \emph
  {et~al.}(2000{\natexlab{c}})\citenamefont {Barnett}, \citenamefont {Pegg},
  \citenamefont {Jeffers}, \citenamefont {Jedrkiewicz},\ and\ \citenamefont
  {Loudon}}]{Barnett2000c}%
  \BibitemOpen
  \bibfield  {author} {\bibinfo {author} {\bibfnamefont {S.~M.}\ \bibnamefont
  {Barnett}}, \bibinfo {author} {\bibfnamefont {D.~T.}\ \bibnamefont {Pegg}},
  \bibinfo {author} {\bibfnamefont {J.}~\bibnamefont {Jeffers}}, \bibinfo
  {author} {\bibfnamefont {O.}~\bibnamefont {Jedrkiewicz}}, \ and\ \bibinfo
  {author} {\bibfnamefont {R.}~\bibnamefont {Loudon}},\ }\href@noop {}
  {\bibfield  {journal} {\bibinfo  {journal} {Phys. Rev. A}\ }\textbf {\bibinfo
  {volume} {62}},\ \bibinfo {pages} {022313} (\bibinfo {year}
  {2000}{\natexlab{c}})}\BibitemShut {NoStop}%
\bibitem [{\citenamefont {Barnett}\ \emph {et~al.}(2001)\citenamefont
  {Barnett}, \citenamefont {Pegg}, \citenamefont {Jeffers},\ and\ \citenamefont
  {Jedrkiewicz}}]{Barnett2001}%
  \BibitemOpen
  \bibfield  {author} {\bibinfo {author} {\bibfnamefont {S.~M.}\ \bibnamefont
  {Barnett}}, \bibinfo {author} {\bibfnamefont {D.~T.}\ \bibnamefont {Pegg}},
  \bibinfo {author} {\bibfnamefont {J.}~\bibnamefont {Jeffers}}, \ and\
  \bibinfo {author} {\bibfnamefont {O.}~\bibnamefont {Jedrkiewicz}},\
  }\href@noop {} {\bibfield  {journal} {\bibinfo  {journal} {Phys. Rev. Lett.}\
  }\textbf {\bibinfo {volume} {86}},\ \bibinfo {pages} {2455} (\bibinfo {year}
  {2001})}\BibitemShut {NoStop}%
\bibitem [{\citenamefont {Jeffers}\ \emph
  {et~al.}(2002{\natexlab{a}})\citenamefont {Jeffers}, \citenamefont
  {Barnett},\ and\ \citenamefont {Pegg}}]{Jeffers2002a}%
  \BibitemOpen
  \bibfield  {author} {\bibinfo {author} {\bibfnamefont {J.}~\bibnamefont
  {Jeffers}}, \bibinfo {author} {\bibfnamefont {S.~M.}\ \bibnamefont
  {Barnett}}, \ and\ \bibinfo {author} {\bibfnamefont {D.~T.}\ \bibnamefont
  {Pegg}},\ }\href@noop {} {\bibfield  {journal} {\bibinfo  {journal} {J. Mod.
  Opt.}\ }\textbf {\bibinfo {volume} {49}},\ \bibinfo {pages} {1175} (\bibinfo
  {year} {2002}{\natexlab{a}})}\BibitemShut {NoStop}%
\bibitem [{\citenamefont {Pregnell}\ and\ \citenamefont
  {Pegg}(2001)}]{Pregnell2001}%
  \BibitemOpen
  \bibfield  {author} {\bibinfo {author} {\bibfnamefont {K.~T.}\ \bibnamefont
  {Pregnell}}\ and\ \bibinfo {author} {\bibfnamefont {D.~T.}\ \bibnamefont
  {Pegg}},\ }\href@noop {} {\bibfield  {journal} {\bibinfo  {journal} {J. Mod.
  Opt.}\ }\textbf {\bibinfo {volume} {48}},\ \bibinfo {pages} {1293} (\bibinfo
  {year} {2001})}\BibitemShut {NoStop}%
\bibitem [{\citenamefont {Pegg}\ and\ \citenamefont
  {Jeffers}(2005)}]{Pegg2005}%
  \BibitemOpen
  \bibfield  {author} {\bibinfo {author} {\bibfnamefont {D.~T.}\ \bibnamefont
  {Pegg}}\ and\ \bibinfo {author} {\bibfnamefont {J.}~\bibnamefont {Jeffers}},\
  }\href@noop {} {\bibfield  {journal} {\bibinfo  {journal} {J. Mod. Opt.}\
  }\textbf {\bibinfo {volume} {52}},\ \bibinfo {pages} {1835} (\bibinfo {year}
  {2005})}\BibitemShut {NoStop}%
\bibitem [{\citenamefont {Resch}\ \emph {et~al.}(2007)\citenamefont {Resch},
  \citenamefont {Pregnell}, \citenamefont {Prevedel}, \citenamefont
  {Gilchrist}, \citenamefont {Pryde}, \citenamefont {O'Brien},\ and\
  \citenamefont {White}}]{Resch2007}%
  \BibitemOpen
  \bibfield  {author} {\bibinfo {author} {\bibfnamefont {K.~J.}\ \bibnamefont
  {Resch}}, \bibinfo {author} {\bibfnamefont {K.~L.}\ \bibnamefont {Pregnell}},
  \bibinfo {author} {\bibfnamefont {R.}~\bibnamefont {Prevedel}}, \bibinfo
  {author} {\bibfnamefont {A.}~\bibnamefont {Gilchrist}}, \bibinfo {author}
  {\bibfnamefont {G.~J.}\ \bibnamefont {Pryde}}, \bibinfo {author}
  {\bibfnamefont {J.~L.}\ \bibnamefont {O'Brien}}, \ and\ \bibinfo {author}
  {\bibfnamefont {A.~G.}\ \bibnamefont {White}},\ }\href@noop {} {\bibfield
  {journal} {\bibinfo  {journal} {Phys. Rev. Lett.}\ }\textbf {\bibinfo
  {volume} {98}},\ \bibinfo {pages} {223601} (\bibinfo {year}
  {2007})}\BibitemShut {NoStop}%
\bibitem [{\citenamefont {Jeffers}\ \emph
  {et~al.}(2002{\natexlab{b}})\citenamefont {Jeffers}, \citenamefont
  {Barnett},\ and\ \citenamefont {Pegg}}]{Jeffers2002b}%
  \BibitemOpen
  \bibfield  {author} {\bibinfo {author} {\bibfnamefont {J.}~\bibnamefont
  {Jeffers}}, \bibinfo {author} {\bibfnamefont {S.~M.}\ \bibnamefont
  {Barnett}}, \ and\ \bibinfo {author} {\bibfnamefont {D.~T.}\ \bibnamefont
  {Pegg}},\ }\href@noop {} {\bibfield  {journal} {\bibinfo  {journal} {J. Mod.
  Opt.}\ }\textbf {\bibinfo {volume} {49}},\ \bibinfo {pages} {925} (\bibinfo
  {year} {2002}{\natexlab{b}})}\BibitemShut {NoStop}%
\bibitem [{\citenamefont {Tan}\ \emph {et~al.}(2004)\citenamefont {Tan},
  \citenamefont {Jeffers},\ and\ \citenamefont {Barnett}}]{Tan2004}%
  \BibitemOpen
  \bibfield  {author} {\bibinfo {author} {\bibfnamefont {E.-K.}\ \bibnamefont
  {Tan}}, \bibinfo {author} {\bibfnamefont {J.}~\bibnamefont {Jeffers}}, \ and\
  \bibinfo {author} {\bibfnamefont {S.~M.}\ \bibnamefont {Barnett}},\
  }\href@noop {} {\bibfield  {journal} {\bibinfo  {journal} {Phys. Rev. A}\
  }\textbf {\bibinfo {volume} {69}},\ \bibinfo {pages} {043806} (\bibinfo
  {year} {2004})}\BibitemShut {NoStop}%
\bibitem [{\citenamefont {Ban}(2007)}]{Ban2007}%
  \BibitemOpen
  \bibfield  {author} {\bibinfo {author} {\bibfnamefont {M.}~\bibnamefont
  {Ban}},\ }\href@noop {} {\bibfield  {journal} {\bibinfo  {journal} {Int. J.
  Theor. Phys.}\ }\textbf {\bibinfo {volume} {46}},\ \bibinfo {pages} {189}
  (\bibinfo {year} {2007})}\BibitemShut {NoStop}%
\bibitem [{\citenamefont {Jedrkiewicz}\ \emph {et~al.}(2004)\citenamefont
  {Jedrkiewicz}, \citenamefont {Loudon},\ and\ \citenamefont
  {Jeffers}}]{Jedrkiewicz2004}%
  \BibitemOpen
  \bibfield  {author} {\bibinfo {author} {\bibfnamefont {O.}~\bibnamefont
  {Jedrkiewicz}}, \bibinfo {author} {\bibfnamefont {R.}~\bibnamefont {Loudon}},
  \ and\ \bibinfo {author} {\bibfnamefont {J.}~\bibnamefont {Jeffers}},\
  }\href@noop {} {\bibfield  {journal} {\bibinfo  {journal} {Phys. Rev. A}\
  }\textbf {\bibinfo {volume} {70}},\ \bibinfo {pages} {033805} (\bibinfo
  {year} {2004})}\BibitemShut {NoStop}%
\bibitem [{\citenamefont {Korotkov}(1999)}]{Korotkov1999}%
  \BibitemOpen
  \bibfield  {author} {\bibinfo {author} {\bibfnamefont {A.~N.}\ \bibnamefont
  {Korotkov}},\ }\href@noop {} {\bibfield  {journal} {\bibinfo  {journal}
  {Phys. Rev. B}\ }\textbf {\bibinfo {volume} {60}},\ \bibinfo {pages} {5737}
  (\bibinfo {year} {1999})}\BibitemShut {NoStop}%
\bibitem [{\citenamefont {Korotkov}\ and\ \citenamefont
  {Jordan}(2006)}]{Korotkov2006}%
  \BibitemOpen
  \bibfield  {author} {\bibinfo {author} {\bibfnamefont {A.~N.}\ \bibnamefont
  {Korotkov}}\ and\ \bibinfo {author} {\bibfnamefont {A.~N.}\ \bibnamefont
  {Jordan}},\ }\href@noop {} {\bibfield  {journal} {\bibinfo  {journal} {Phys.
  Rev. Lett.}\ }\textbf {\bibinfo {volume} {97}},\ \bibinfo {pages} {166805}
  (\bibinfo {year} {2006})}\BibitemShut {NoStop}%
\bibitem [{\citenamefont {Jordan}\ and\ \citenamefont
  {Korotkov}(2006)}]{Jordan2006}%
  \BibitemOpen
  \bibfield  {author} {\bibinfo {author} {\bibfnamefont {A.~N.}\ \bibnamefont
  {Jordan}}\ and\ \bibinfo {author} {\bibfnamefont {A.~N.}\ \bibnamefont
  {Korotkov}},\ }\href@noop {} {\bibfield  {journal} {\bibinfo  {journal}
  {Phys. Rev. B}\ }\textbf {\bibinfo {volume} {74}},\ \bibinfo {pages} {085307}
  (\bibinfo {year} {2006})}\BibitemShut {NoStop}%
\bibitem [{\citenamefont {Williams}\ and\ \citenamefont
  {Jordan}(2008)}]{Williams2008}%
  \BibitemOpen
  \bibfield  {author} {\bibinfo {author} {\bibfnamefont {N.~S.}\ \bibnamefont
  {Williams}}\ and\ \bibinfo {author} {\bibfnamefont {A.~N.}\ \bibnamefont
  {Jordan}},\ }\href@noop {} {\bibfield  {journal} {\bibinfo  {journal} {Phys.
  Rev. Lett.}\ }\textbf {\bibinfo {volume} {100}},\ \bibinfo {pages} {026804}
  (\bibinfo {year} {2008})}\BibitemShut {NoStop}%
\bibitem [{\citenamefont {Caves}\ \emph {et~al.}(2002)\citenamefont {Caves},
  \citenamefont {Fuchs},\ and\ \citenamefont {Schack}}]{Caves2002}%
  \BibitemOpen
  \bibfield  {author} {\bibinfo {author} {\bibfnamefont {C.~M.}\ \bibnamefont
  {Caves}}, \bibinfo {author} {\bibfnamefont {C.~A.}\ \bibnamefont {Fuchs}}, \
  and\ \bibinfo {author} {\bibfnamefont {R.}~\bibnamefont {Schack}},\
  }\href@noop {} {\bibfield  {journal} {\bibinfo  {journal} {Phys. Rev. A}\
  }\textbf {\bibinfo {volume} {65}},\ \bibinfo {pages} {22305} (\bibinfo {year}
  {2002})}\BibitemShut {NoStop}%
\bibitem [{\citenamefont {Spekkens}(2007)}]{Spekkens2007}%
  \BibitemOpen
  \bibfield  {author} {\bibinfo {author} {\bibfnamefont {R.~W.}\ \bibnamefont
  {Spekkens}},\ }\href@noop {} {\bibfield  {journal} {\bibinfo  {journal}
  {Phys. Rev. A}\ }\textbf {\bibinfo {volume} {75}},\ \bibinfo {pages} {032110}
  (\bibinfo {year} {2007})}\BibitemShut {NoStop}%
\bibitem [{\citenamefont {Harrigan}\ and\ \citenamefont
  {Spekkens}(2010)}]{Harrigan2010}%
  \BibitemOpen
  \bibfield  {author} {\bibinfo {author} {\bibfnamefont {N.}~\bibnamefont
  {Harrigan}}\ and\ \bibinfo {author} {\bibfnamefont {R.~W.}\ \bibnamefont
  {Spekkens}},\ }\href@noop {} {\bibfield  {journal} {\bibinfo  {journal}
  {Found. Phys.}\ }\textbf {\bibinfo {volume} {40}},\ \bibinfo {pages} {125}
  (\bibinfo {year} {2010})}\BibitemShut {NoStop}%
\bibitem [{\citenamefont {Bartlett}\ \emph {et~al.}(2012)\citenamefont
  {Bartlett}, \citenamefont {Rudolph},\ and\ \citenamefont
  {Spekkens}}]{Bartlett2012}%
  \BibitemOpen
  \bibfield  {author} {\bibinfo {author} {\bibfnamefont {S.~D.}\ \bibnamefont
  {Bartlett}}, \bibinfo {author} {\bibfnamefont {T.}~\bibnamefont {Rudolph}}, \
  and\ \bibinfo {author} {\bibfnamefont {R.~W.}\ \bibnamefont {Spekkens}},\
  }\href@noop {} {\bibfield  {journal} {\bibinfo  {journal} {Phys. Rev. A}\
  }\textbf {\bibinfo {volume} {86}},\ \bibinfo {pages} {012103} (\bibinfo
  {year} {2012})}\BibitemShut {NoStop}%
\bibitem [{\citenamefont {Leifer}(2006)}]{Leifer2006}%
  \BibitemOpen
  \bibfield  {author} {\bibinfo {author} {\bibfnamefont {M.~S.}\ \bibnamefont
  {Leifer}},\ }\href@noop {} {\bibfield  {journal} {\bibinfo  {journal} {Phys.
  Rev. A}\ }\textbf {\bibinfo {volume} {74}},\ \bibinfo {pages} {042310}
  (\bibinfo {year} {2006})}\BibitemShut {NoStop}%
\bibitem [{\citenamefont {Leifer}(2007)}]{Leifer2007}%
  \BibitemOpen
  \bibfield  {author} {\bibinfo {author} {\bibfnamefont {M.~S.}\ \bibnamefont
  {Leifer}},\ }in\ \href@noop {} {\emph {\bibinfo {booktitle} {{AIP Conference
  Proceedings vol. 889}}}},\ \bibinfo {editor} {edited by\ \bibinfo {editor}
  {\bibfnamefont {G.}~\bibnamefont {Adenier}}, \bibinfo {editor} {\bibfnamefont
  {C.~A.}\ \bibnamefont {Fuchs}}, \ and\ \bibinfo {editor} {\bibfnamefont
  {A.~Y.}\ \bibnamefont {Khrennikov}}}\ (\bibinfo {year} {2007})\ pp.\ \bibinfo
  {pages} {172--186}\BibitemShut {NoStop}%
\bibitem [{\citenamefont {Leifer}\ and\ \citenamefont
  {Poulin}(2008)}]{Leifer2008}%
  \BibitemOpen
  \bibfield  {author} {\bibinfo {author} {\bibfnamefont {M.~S.}\ \bibnamefont
  {Leifer}}\ and\ \bibinfo {author} {\bibfnamefont {D.}~\bibnamefont
  {Poulin}},\ }\href@noop {} {\bibfield  {journal} {\bibinfo  {journal} {Ann.
  Phys.}\ }\textbf {\bibinfo {volume} {323}},\ \bibinfo {pages} {1899}
  (\bibinfo {year} {2008})}\BibitemShut {NoStop}%
\bibitem [{\citenamefont {Leifer}\ and\ \citenamefont
  {Spekkens}(2011)}]{Leifer2011}%
  \BibitemOpen
  \bibfield  {author} {\bibinfo {author} {\bibfnamefont {M.~S.}\ \bibnamefont
  {Leifer}}\ and\ \bibinfo {author} {\bibfnamefont {R.~W.}\ \bibnamefont
  {Spekkens}},\ }\href@noop {} {\enquote {\bibinfo {title} {{Formulating
  Quantum Theory as a Causally Neutral Theory of Bayesian Inference}},}\ }
  (\bibinfo {year} {2011}),\ \Eprint {http://arxiv.org/abs/arXiv:1107.5849}
  {arXiv:1107.5849} \BibitemShut {NoStop}%
\bibitem [{\citenamefont {Abramsky}\ and\ \citenamefont
  {Coecke}(2004)}]{Abramsky2004}%
  \BibitemOpen
  \bibfield  {author} {\bibinfo {author} {\bibfnamefont {S.}~\bibnamefont
  {Abramsky}}\ and\ \bibinfo {author} {\bibfnamefont {B.}~\bibnamefont
  {Coecke}},\ }in\ \href@noop {} {\emph {\bibinfo {booktitle} {{Proceedings of
  19th IEEE conference on Logic in Computer Science}}}}\ (\bibinfo  {publisher}
  {IEEE Press Washington, DC},\ \bibinfo {year} {2004})\ pp.\ \bibinfo {pages}
  {415--425}\BibitemShut {NoStop}%
\bibitem [{\citenamefont {Selinger}(2004)}]{Selinger2004}%
  \BibitemOpen
  \bibfield  {author} {\bibinfo {author} {\bibfnamefont {P.}~\bibnamefont
  {Selinger}},\ }\href@noop {} {\bibfield  {journal} {\bibinfo  {journal}
  {Math. Struct. in Comp. Science}\ }\textbf {\bibinfo {volume} {14}},\
  \bibinfo {pages} {527} (\bibinfo {year} {2004})}\BibitemShut {NoStop}%
\bibitem [{\citenamefont {Selinger}(2007)}]{Selinger2007}%
  \BibitemOpen
  \bibfield  {author} {\bibinfo {author} {\bibfnamefont {P.}~\bibnamefont
  {Selinger}},\ }\href@noop {} {\bibfield  {journal} {\bibinfo  {journal}
  {Electronic Notes in Theoretical Computer Science}\ }\textbf {\bibinfo
  {volume} {170}},\ \bibinfo {pages} {139} (\bibinfo {year}
  {2007})}\BibitemShut {NoStop}%
\bibitem [{\citenamefont {Coecke}\ and\ \citenamefont
  {Spekkens}(2012)}]{Coecke2012}%
  \BibitemOpen
  \bibfield  {author} {\bibinfo {author} {\bibfnamefont {B.}~\bibnamefont
  {Coecke}}\ and\ \bibinfo {author} {\bibfnamefont {R.~W.}\ \bibnamefont
  {Spekkens}},\ }\href@noop {} {\bibfield  {journal} {\bibinfo  {journal}
  {Synthese}\ }\textbf {\bibinfo {volume} {186}},\ \bibinfo {pages} {651}
  (\bibinfo {year} {2012})}\BibitemShut {NoStop}%
\bibitem [{\citenamefont {Rau}(2011)}]{Rau2011}%
  \BibitemOpen
  \bibfield  {author} {\bibinfo {author} {\bibfnamefont {J.}~\bibnamefont
  {Rau}},\ }\href@noop {} {\bibfield  {journal} {\bibinfo  {journal} {Found.
  Phys.}\ }\textbf {\bibinfo {volume} {41}},\ \bibinfo {pages} {380} (\bibinfo
  {year} {2011})}\BibitemShut {NoStop}%
\bibitem [{\citenamefont {Davies}\ and\ \citenamefont
  {Lewis}(1970)}]{Davies1970}%
  \BibitemOpen
  \bibfield  {author} {\bibinfo {author} {\bibfnamefont {E.~B.}\ \bibnamefont
  {Davies}}\ and\ \bibinfo {author} {\bibfnamefont {J.~T.}\ \bibnamefont
  {Lewis}},\ }\href@noop {} {\bibfield  {journal} {\bibinfo  {journal} {Comm.
  Math. Phys.}\ }\textbf {\bibinfo {volume} {17}},\ \bibinfo {pages} {239}
  (\bibinfo {year} {1970})}\BibitemShut {NoStop}%
\bibitem [{\citenamefont {Ozawa}(1984)}]{Ozawa1984}%
  \BibitemOpen
  \bibfield  {author} {\bibinfo {author} {\bibfnamefont {M.}~\bibnamefont
  {Ozawa}},\ }\href@noop {} {\bibfield  {journal} {\bibinfo  {journal} {J.
  Math. Phys.}\ }\textbf {\bibinfo {volume} {25}},\ \bibinfo {pages} {79}
  (\bibinfo {year} {1984})}\BibitemShut {NoStop}%
\bibitem [{\citenamefont {Aharonov}\ \emph {et~al.}(1988)\citenamefont
  {Aharonov}, \citenamefont {Albert},\ and\ \citenamefont
  {Vaidman}}]{Aharonov1988}%
  \BibitemOpen
  \bibfield  {author} {\bibinfo {author} {\bibfnamefont {Y.}~\bibnamefont
  {Aharonov}}, \bibinfo {author} {\bibfnamefont {D.~Z.}\ \bibnamefont
  {Albert}}, \ and\ \bibinfo {author} {\bibfnamefont {L.}~\bibnamefont
  {Vaidman}},\ }\href@noop {} {\bibfield  {journal} {\bibinfo  {journal} {Phys.
  Rev. Lett.}\ }\textbf {\bibinfo {volume} {60}},\ \bibinfo {pages} {1351 }
  (\bibinfo {year} {1988})}\BibitemShut {NoStop}%
\bibitem [{\citenamefont {Duck}\ \emph {et~al.}(1989)\citenamefont {Duck},
  \citenamefont {Stevenson},\ and\ \citenamefont {Sudarshan}}]{Duck1989}%
  \BibitemOpen
  \bibfield  {author} {\bibinfo {author} {\bibfnamefont {I.~M.}\ \bibnamefont
  {Duck}}, \bibinfo {author} {\bibfnamefont {P.~M.}\ \bibnamefont {Stevenson}},
  \ and\ \bibinfo {author} {\bibfnamefont {E.~C.~G.}\ \bibnamefont
  {Sudarshan}},\ }\href@noop {} {\bibfield  {journal} {\bibinfo  {journal}
  {Phys. Rev. D}\ }\textbf {\bibinfo {volume} {40}},\ \bibinfo {pages} {2112}
  (\bibinfo {year} {1989})}\BibitemShut {NoStop}%
\bibitem [{\citenamefont {Aharonov}\ and\ \citenamefont
  {Vaidman}(2008)}]{Aharonov2008}%
  \BibitemOpen
  \bibfield  {author} {\bibinfo {author} {\bibfnamefont {Y.}~\bibnamefont
  {Aharonov}}\ and\ \bibinfo {author} {\bibfnamefont {L.}~\bibnamefont
  {Vaidman}},\ }\href@noop {} {\bibfield  {journal} {\bibinfo  {journal} {Lect.
  Notes Phys.}\ }\textbf {\bibinfo {volume} {734}},\ \bibinfo {pages} {399}
  (\bibinfo {year} {2008})}\BibitemShut {NoStop}%
\bibitem [{\citenamefont {Aharonov}\ \emph {et~al.}(2009)\citenamefont
  {Aharonov}, \citenamefont {Popescu}, \citenamefont {Tollaksen},\ and\
  \citenamefont {Vaidman}}]{Aharonov2009}%
  \BibitemOpen
  \bibfield  {author} {\bibinfo {author} {\bibfnamefont {Y.}~\bibnamefont
  {Aharonov}}, \bibinfo {author} {\bibfnamefont {S.}~\bibnamefont {Popescu}},
  \bibinfo {author} {\bibfnamefont {J.}~\bibnamefont {Tollaksen}}, \ and\
  \bibinfo {author} {\bibfnamefont {L.}~\bibnamefont {Vaidman}},\ }\href@noop
  {} {\bibfield  {journal} {\bibinfo  {journal} {Phys. Rev. A}\ }\textbf
  {\bibinfo {volume} {79}},\ \bibinfo {pages} {052110} (\bibinfo {year}
  {2009})}\BibitemShut {NoStop}%
\bibitem [{\citenamefont {Aharonov}\ \emph {et~al.}(2010)\citenamefont
  {Aharonov}, \citenamefont {Popescu},\ and\ \citenamefont
  {Tollaksen}}]{Aharonov2010}%
  \BibitemOpen
  \bibfield  {author} {\bibinfo {author} {\bibfnamefont {Y.}~\bibnamefont
  {Aharonov}}, \bibinfo {author} {\bibfnamefont {S.}~\bibnamefont {Popescu}}, \
  and\ \bibinfo {author} {\bibfnamefont {J.}~\bibnamefont {Tollaksen}},\
  }\href@noop {} {\bibfield  {journal} {\bibinfo  {journal} {Physics Today}\
  }\textbf {\bibinfo {volume} {63}},\ \bibinfo {pages} {27} (\bibinfo {year}
  {2010})}\BibitemShut {NoStop}%
\bibitem [{\citenamefont {Crutchfield}\ \emph {et~al.}(2009)\citenamefont
  {Crutchfield}, \citenamefont {Ellison},\ and\ \citenamefont
  {Maloney}}]{Crutchfield2009}%
  \BibitemOpen
  \bibfield  {author} {\bibinfo {author} {\bibfnamefont {J.~P.}\ \bibnamefont
  {Crutchfield}}, \bibinfo {author} {\bibfnamefont {C.~J.}\ \bibnamefont
  {Ellison}}, \ and\ \bibinfo {author} {\bibfnamefont {J.~R.}\ \bibnamefont
  {Maloney}},\ }\href@noop {} {\bibfield  {journal} {\bibinfo  {journal} {Phys.
  Rev. Lett.}\ }\textbf {\bibinfo {volume} {103}},\ \bibinfo {pages} {094101}
  (\bibinfo {year} {2009})}\BibitemShut {NoStop}%
\bibitem [{\citenamefont {Ellison}\ \emph {et~al.}(2009)\citenamefont
  {Ellison}, \citenamefont {Mahoney},\ and\ \citenamefont
  {Crutchfield}}]{Ellison2009}%
  \BibitemOpen
  \bibfield  {author} {\bibinfo {author} {\bibfnamefont {C.~J.}\ \bibnamefont
  {Ellison}}, \bibinfo {author} {\bibfnamefont {J.~R.}\ \bibnamefont
  {Mahoney}}, \ and\ \bibinfo {author} {\bibfnamefont {J.~P.}\ \bibnamefont
  {Crutchfield}},\ }\href@noop {} {\bibfield  {journal} {\bibinfo  {journal}
  {J. Stat. Phys.}\ }\textbf {\bibinfo {volume} {136}},\ \bibinfo {pages}
  {1005} (\bibinfo {year} {2009})}\BibitemShut {NoStop}%
\bibitem [{\citenamefont {Ellison}\ \emph {et~al.}(2011)\citenamefont
  {Ellison}, \citenamefont {Mahoney}, \citenamefont {James}, \citenamefont
  {Crutchfield},\ and\ \citenamefont {Reichardt}}]{Ellison2011}%
  \BibitemOpen
  \bibfield  {author} {\bibinfo {author} {\bibfnamefont {C.~J.}\ \bibnamefont
  {Ellison}}, \bibinfo {author} {\bibfnamefont {J.~R.}\ \bibnamefont
  {Mahoney}}, \bibinfo {author} {\bibfnamefont {R.~G.}\ \bibnamefont {James}},
  \bibinfo {author} {\bibfnamefont {J.~P.}\ \bibnamefont {Crutchfield}}, \ and\
  \bibinfo {author} {\bibfnamefont {J.}~\bibnamefont {Reichardt}},\ }\href@noop
  {} {\bibfield  {journal} {\bibinfo  {journal} {Chaos}\ }\textbf {\bibinfo
  {volume} {21}},\ \bibinfo {pages} {037107} (\bibinfo {year}
  {2011})}\BibitemShut {NoStop}%
\bibitem [{\citenamefont {Nielsen}\ and\ \citenamefont
  {Chuang}(2000)}]{Nielsen2000}%
  \BibitemOpen
  \bibfield  {author} {\bibinfo {author} {\bibfnamefont {M.~A.}\ \bibnamefont
  {Nielsen}}\ and\ \bibinfo {author} {\bibfnamefont {I.~L.}\ \bibnamefont
  {Chuang}},\ }\href@noop {} {\emph {\bibinfo {title} {Quantum computation and
  quantum information}}}\ (\bibinfo  {publisher} {Cambridge University Press,
  Cambridge},\ \bibinfo {year} {2000})\BibitemShut {NoStop}%
\bibitem [{\citenamefont {Kraus}(1971)}]{Kraus1971}%
  \BibitemOpen
  \bibfield  {author} {\bibinfo {author} {\bibfnamefont {K.}~\bibnamefont
  {Kraus}},\ }\href@noop {} {\bibfield  {journal} {\bibinfo  {journal} {Annals
  of Physics}\ }\textbf {\bibinfo {volume} {64}},\ \bibinfo {pages} {311}
  (\bibinfo {year} {1971})}\BibitemShut {NoStop}%
\bibitem [{\citenamefont {Gleason}(1957)}]{Gleason1957}%
  \BibitemOpen
  \bibfield  {author} {\bibinfo {author} {\bibfnamefont {A.}~\bibnamefont
  {Gleason}},\ }\href@noop {} {\bibfield  {journal} {\bibinfo  {journal}
  {Indiana Univ. Math. J.}\ }\textbf {\bibinfo {volume} {6}},\ \bibinfo {pages}
  {885} (\bibinfo {year} {1957})}\BibitemShut {NoStop}%
\bibitem [{Note1()}]{Note1}%
  \BibitemOpen
  \bibinfo {note} {A POM also has the common name of positive-operator-valued
  measure (POVM).}\BibitemShut {Stop}%
\bibitem [{\citenamefont {Dressel}\ \emph {et~al.}(2010)\citenamefont
  {Dressel}, \citenamefont {Agarwal},\ and\ \citenamefont
  {Jordan}}]{Dressel2010}%
  \BibitemOpen
  \bibfield  {author} {\bibinfo {author} {\bibfnamefont {J.}~\bibnamefont
  {Dressel}}, \bibinfo {author} {\bibfnamefont {S.}~\bibnamefont {Agarwal}}, \
  and\ \bibinfo {author} {\bibfnamefont {A.~N.}\ \bibnamefont {Jordan}},\
  }\href@noop {} {\bibfield  {journal} {\bibinfo  {journal} {Phys. Rev. Lett.}\
  }\textbf {\bibinfo {volume} {104}},\ \bibinfo {pages} {240401} (\bibinfo
  {year} {2010})}\BibitemShut {NoStop}%
\bibitem [{\citenamefont {Robinson}(1966)}]{Robinson1966}%
  \BibitemOpen
  \bibfield  {author} {\bibinfo {author} {\bibfnamefont {A.}~\bibnamefont
  {Robinson}},\ }\href@noop {} {\emph {\bibinfo {title} {{Non-standard
  Analysis}}}}\ (\bibinfo  {publisher} {Princeton University Press},\ \bibinfo
  {year} {1966})\BibitemShut {NoStop}%
\bibitem [{Note2()}]{Note2}%
  \BibitemOpen
  \bibinfo {note} {Such an infinite constant can be given rigorous meaning as
  an element of an expanded number field, such as Robinson's non-standard
  integers \cite {Robinson1966}.}\BibitemShut {Stop}%
\bibitem [{\citenamefont {Breuer}\ and\ \citenamefont
  {Petruccione}(2007)}]{Breuer2007}%
  \BibitemOpen
  \bibfield  {author} {\bibinfo {author} {\bibfnamefont {H.}~\bibnamefont
  {Breuer}}\ and\ \bibinfo {author} {\bibfnamefont {F.}~\bibnamefont
  {Petruccione}},\ }\href@noop {} {\emph {\bibinfo {title} {The Theory of Open
  Quantum Systems}}}\ (\bibinfo  {publisher} {Oxford University Press,
  Oxford},\ \bibinfo {year} {2007})\BibitemShut {NoStop}%
\bibitem [{\citenamefont {Dressel}\ \emph {et~al.}(2011)\citenamefont
  {Dressel}, \citenamefont {Broadbent}, \citenamefont {Howell},\ and\
  \citenamefont {Jordan}}]{Dressel2011}%
  \BibitemOpen
  \bibfield  {author} {\bibinfo {author} {\bibfnamefont {J.}~\bibnamefont
  {Dressel}}, \bibinfo {author} {\bibfnamefont {C.~J.}\ \bibnamefont
  {Broadbent}}, \bibinfo {author} {\bibfnamefont {J.~C.}\ \bibnamefont
  {Howell}}, \ and\ \bibinfo {author} {\bibfnamefont {A.~N.}\ \bibnamefont
  {Jordan}},\ }\href@noop {} {\bibfield  {journal} {\bibinfo  {journal} {Phys.
  Rev. Lett.}\ }\textbf {\bibinfo {volume} {106}},\ \bibinfo {pages} {040402}
  (\bibinfo {year} {2011})}\BibitemShut {NoStop}%
\bibitem [{\citenamefont {Dressel}\ and\ \citenamefont
  {Jordan}(2012{\natexlab{b}})}]{Dressel2012}%
  \BibitemOpen
  \bibfield  {author} {\bibinfo {author} {\bibfnamefont {J.}~\bibnamefont
  {Dressel}}\ and\ \bibinfo {author} {\bibfnamefont {A.~N.}\ \bibnamefont
  {Jordan}},\ }\href@noop {} {\bibfield  {journal} {\bibinfo  {journal} {J.
  Phys. A: Math. Theor.}\ }\textbf {\bibinfo {volume} {45}},\ \bibinfo {pages}
  {015304} (\bibinfo {year} {2012}{\natexlab{b}})}\BibitemShut {NoStop}%
\bibitem [{\citenamefont {Dressel}\ \emph {et~al.}(2012)\citenamefont
  {Dressel}, \citenamefont {Choi},\ and\ \citenamefont
  {Jordan}}]{Dressel2012c}%
  \BibitemOpen
  \bibfield  {author} {\bibinfo {author} {\bibfnamefont {J.}~\bibnamefont
  {Dressel}}, \bibinfo {author} {\bibfnamefont {Y.}~\bibnamefont {Choi}}, \
  and\ \bibinfo {author} {\bibfnamefont {A.~N.}\ \bibnamefont {Jordan}},\
  }\href@noop {} {\bibfield  {journal} {\bibinfo  {journal} {Phys. Rev. B}\
  }\textbf {\bibinfo {volume} {85}},\ \bibinfo {pages} {045320} (\bibinfo
  {year} {2012})}\BibitemShut {NoStop}%
\bibitem [{\citenamefont {Dressel}\ and\ \citenamefont
  {Jordan}(2012{\natexlab{c}})}]{Dressel2012d}%
  \BibitemOpen
  \bibfield  {author} {\bibinfo {author} {\bibfnamefont {J.}~\bibnamefont
  {Dressel}}\ and\ \bibinfo {author} {\bibfnamefont {A.~N.}\ \bibnamefont
  {Jordan}},\ }\href@noop {} {\bibfield  {journal} {\bibinfo  {journal} {Phys.
  Rev. A}\ }\textbf {\bibinfo {volume} {85}},\ \bibinfo {pages} {012107}
  (\bibinfo {year} {2012}{\natexlab{c}})}\BibitemShut {NoStop}%
\bibitem [{\citenamefont {Dressel}\ and\ \citenamefont
  {Jordan}(2012{\natexlab{d}})}]{Dressel2012e}%
  \BibitemOpen
  \bibfield  {author} {\bibinfo {author} {\bibfnamefont {J.}~\bibnamefont
  {Dressel}}\ and\ \bibinfo {author} {\bibfnamefont {A.~N.}\ \bibnamefont
  {Jordan}},\ }\href@noop {} {\bibfield  {journal} {\bibinfo  {journal} {Phys.
  Rev. Lett.}\ }\textbf {\bibinfo {volume} {109}},\ \bibinfo {pages} {230402}
  (\bibinfo {year} {2012}{\natexlab{d}})}\BibitemShut {NoStop}%
\end{thebibliography}
%
\end{document}